\documentclass[3p,onecolumn,10pt,times, sort]{elsarticle}

\usepackage{etoolbox}
\usepackage{lineno}
\usepackage{xspace} % to get the spacing after macros right  
\usepackage[dvipsnames]{xcolor}
\usepackage{amsmath,amsfonts,amssymb}
\usepackage{mathtools}
\usepackage{bm}
\usepackage{siunitx}
\usepackage{textcomp}
\usepackage{miller}
\usepackage{tikz}
\usetikzlibrary{trees,matrix}
\usepackage{forest}
\usetikzlibrary{arrows,shapes,calc,positioning}
\usepackage{subfig}
\usepackage{subeqnarray}
\usepackage{multirow}
\usepackage{tabularx}
\usepackage{booktabs}
\usepackage{diagbox} % for diagonal lines in tables
\usepackage[export]{adjustbox}
\usepackage{graphbox}
\usepackage{overpic} % for superimposing images
\usepackage{enumitem} % for changing lists (especially the vertical spacing)
\usepackage{calc}
\usepackage{todonotes}
\usepackage{lipsum}
\usepackage{color,soul}
\usepackage{graphicx}
\usepackage{caption}
\usepackage{longtable,xtab}
\usepackage{hyperref}
\hypersetup{
     colorlinks   = true,
     allcolors    = NavyBlue,
}
\usepackage[capitalize]{cleveref}
\usepackage{doi}
\graphicspath{{./}{./figures/}{../figures/}}
\DeclareGraphicsExtensions{.pdf,.png,.jpg}

\newcommand{\Tisixtwo}{Ti-6Al-2Sn-4Zr-6Mo}

\newcommand{\ie}{\mbox{i.\hspace{1pt}e.}\xspace}

\newcommand{\N}[1]{\ensuremath{N_\text{#1}}}

\newcommand{\hcp}{\ensuremath{\alpha}\xspace}
\newcommand{\bcc}{\ensuremath{\beta}\xspace}

\newcommand{\crystaxis}[1]{\ensuremath{\langle{#1}\rangle}}

\newcommand{\tnsr}[1]{\ensuremath{\bm{#1}}}
\newcommand{\vctr}[1]{\ensuremath{\bm{#1}}}

\newcommand{\transpose}[1]{{#1}^{\mathrm T}}
\newcommand{\inverse}[1]{{#1}^{-1}}

\newcommand{\burgerslength}{\ensuremath{b}}
\DeclareSIUnit\burgersunit{$\burgerslength$}
\DeclareSIUnit\micron{\micro\metre}

\journal{TBD}

\begin{document}

\begin{frontmatter}


\title{\bcc orientation reconstruction and shear deformation calculation in hcp-bcc-hcp phase transformation 
%and an approach to reconstruct the orientations of bcc parent grains from transformed hcp grains
}

\author[CHEMS,CZII]{Zhuowen Zhao\corref{Cor}}
\author[CHEMS]{Thomas R. Bieler}
\author[CHEMS]{Philip Eisenlohr\corref{Cor}}

%\ead{kevin.zhao@czii.org}
\cortext[Cor]{Correspondence: kevin.zhao@czii.org, eisenlohr@egr.msu.edu}

\address[CHEMS]{
Chemical Engineering and Materials Science,
Michigan State University,
East Lansing, MI 48824, USA
}

\address[CZII]{
Chan Zuckerberg Institute for Advanced Biological Imaging (CZ Imaging Institute), Redwood City, CA 94065
}

%% Abstract
\begin{abstract}
We introduce a cluster-based technique to automate pixel-wise reconstruction of \bcc orientations from parent \hcp orientations over large, indexed regions.
This approach provides a valuable tool for analyzing problems that require historical information about current \hcp microstructures, such as investigating variant selection mechanisms during the \(\hcp\to\bcc\to\hcp\) transformation.
Additionally, we present a method for calculating deformation gradient variants associated with phase transformations between hcp (\hcp) and bcc (\bcc) phases based upon a series of frame rotations and shape transformations. 
This method streamlines the integration of transformation kinematics into continuum-based models by enabling convenient computation of the deformation gradient governing the transformation.

\end{abstract}

%% Keywords


\end{frontmatter}

\section{Introduction}
 The bcc-hcp phase transformation is a significant topic for Ti and Zr alloys due to its crucial impact on material properties. 
A comprehensive understanding of the kinematics of this transformation is vital for optimizing and innovating material manufacturing processes, particularly in thermo-mechanical applications. 
Two fundamental calculations---orientation determination and deformation gradient computation---are essential for accurately modeling shape changes and orientation evolution during the transformation. 

%Here, we present a method to compute the deformation gradient in a manner consistent with orientation calculations, adopting a framework based on frame rotation and transformation principles.

The orientation relationships during this transformation has been investigated over the years.
%Reconstruction of the \bcc  phase is critical for studying Ti and Zr alloys when historical information about the current microstructures is required. 
Automated tools for generating reconstructed \bcc  orientation maps from outputs of commercial software, such as Electron Backscatter Diffraction (EBSD) mapping, are in high demand. 
Over the past two decades, numerous methods for \bcc reconstruction have been proposed \citep{Birch+Britton2021, Cayron_etal2006, Glavicic_etal2003, Hadke_etal2016, Krishna_etal2010, Nyyssonen_etal2018, Zaitzeff_etal2021}, differing mainly in how they group \hcp grains originating from the same parent \bcc grain. 
These methods can be broadly categorized into two groups:

\subsection{Grouping the reconstructed \bcc orientations}
This category involves grouping \hcp grains that share similar reconstructed \bcc orientations, which exhibit small variations, indicating they originated from the same parent \bcc grain. 
Notable methods in this category include:

\paragraph{The Monte Carlo method} 
This approach randomly selects pairs of points and calculates transformed \bcc orientations by first minimizing the misorientation angle between any two points, followed by a global minimization of the sum of misorientation angles \citep{Glavicic_etal2003}. 
While this method is highly accurate, it is computationally intensive, as the back-transformation calculation---one of the most resource-demanding components---must be performed beforehand.

\paragraph{Summation of mutual misorientation angle (SMMA) method} 
Similar to the Monte Carlo method, the SMMA method identifies parent \bcc orientations by minimizing the sum of misorientation angles between \bcc members in a single step \citep{Tari_etal2013}. 
This technique produces accurate results but remains computationally expensive.

\paragraph{Image segmentation method} 
This method utilizes the Mumford-Shah variational model, originally developed for image segmentation in computer vision, to cluster distinct \bcc  orientations \citep{Zaitzeff_etal2021}. 
Unlike other methods, it does not require the computation of misorientation angles, instead relying on a robust clustering algorithm. 
Due to the clustering algorithm, the method is also applicable to microstructures that do not follow the Burgers orientation relationship.

\subsection{Grouping \hcp orientations from the same parent grain}
The second category of \bcc reconstruction methods involves grouping \hcp orientations originating from the same parent grain. 
Since back-transformation calculations are not required in this category, these methods are generally more computationally efficient than those based on grouping reconstructed \bcc orientations.

\paragraph{Groupoid method (or the triplet method)}  
This approach begins by identifying triplets of \hcp grains (``nucleus'') with low misorientation angles \citep{Cayron_etal2006, Krishna_etal2010}. 
The triplets are then expanded with relaxed tolerances to encompass the entire map. 
While this non-iterative method is fast, it may fail to group all grains, leaving some unclassified.

\paragraph{Clustering method}
This method is a more efficient variation of the SMMA method \citep{Hadke_etal2016}. 
It focuses on clustering \hcp grains rather than reconstructed \bcc orientations to avoid extensive back-transformation calculations. 
However, like the groupoid method, its accuracy is limited; it may fail to group isolated grains within larger grains or small clusters of child grains.
  
\paragraph{Markov Cluster Algorithm (MCL) method}  
The MCL method leverages a computer vision algorithm designed to detect natural groupings in graphs \citep{Birch+Britton2021, Nyyssonen_etal2018}. 
It may introduce inaccuracies by adding parent grain orientations if an excessive number of clusters are detected.\\

As highlighted in \citep{Birch+Britton2021}, \bcc reconstruction inherently involves a trade-off between efficiency and accuracy: methods that group \hcp grains tend to be more efficient but less accurate than those grouping reconstructed \bcc orientations. 
In this paper, we introduce a reconstruction method designed to achieve a balanced compromise between these two factors.

We begin by summarizing the orientation calculations, which serve as the foundation of \bcc reconstruction. 
Building on this, we propose a cluster-averaging approach to reconstruct \bcc orientations from an \hcp orientation map, utilizing a window-searching algorithm. 
This approach establishes unique \bcc clusters, which should be inherently accurate (the second category characteristics).
Additionally, the window-searching strategy enhances efficiency by limiting back-transformation calculations to spatially adjacent \hcp grains likely originating from the same \bcc parent, leveraging characteristics of the first category. 
This method also inherently supports parallel processing, further reducing computational time.

Furthermore, we present a complementary method in \cref{sec: F calculation} to compute the deformation gradient in a manner consistent with the orientation calculations, adopting a robust framework based on frame rotation and transformations. 
Together, these methods provide a comprehensive and efficient approach to analyzing phase transformation kinematics.

\section{Method}
\label{sec: a-b transformation kinematics}

\begin{figure}[h!]
\includegraphics[width=1\linewidth]{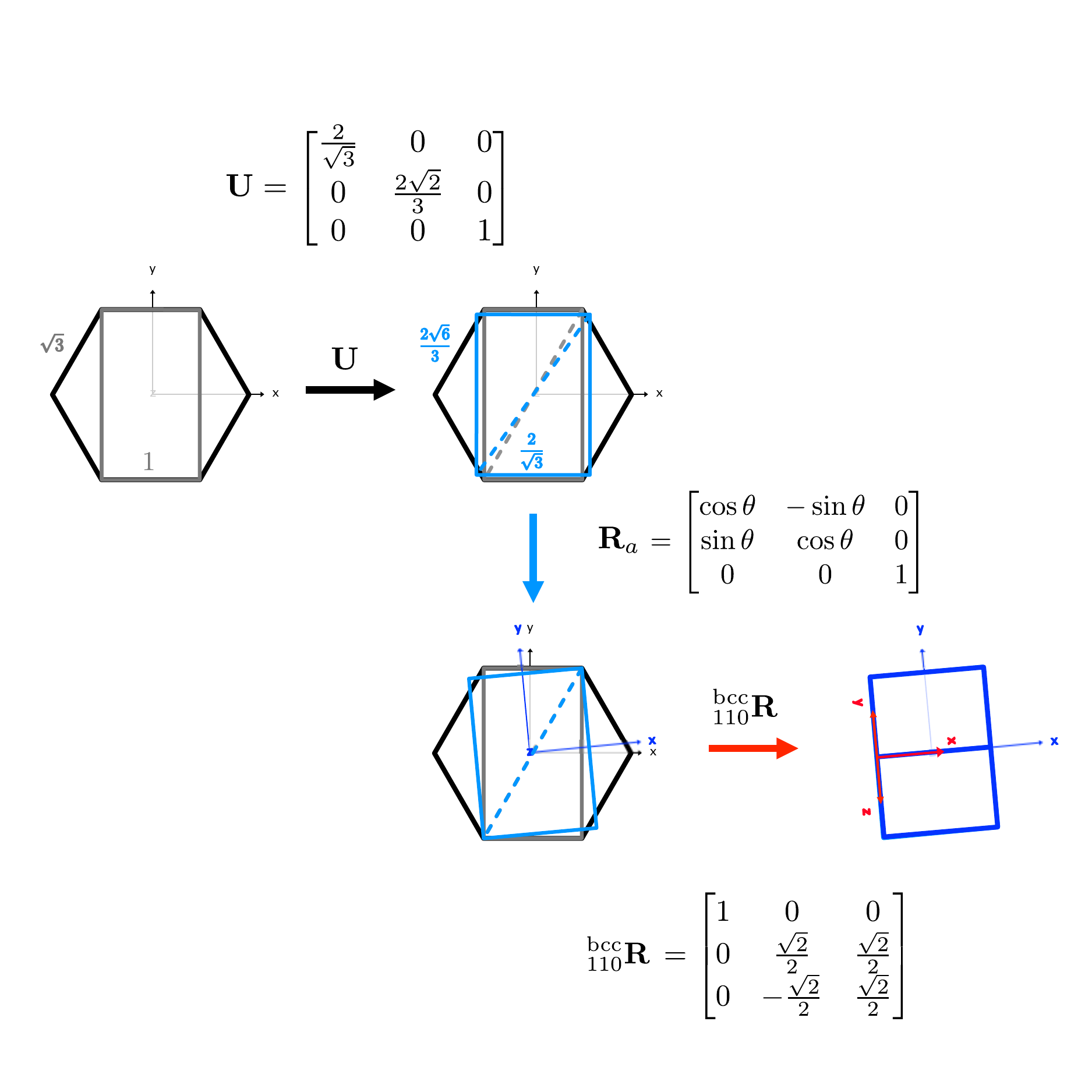}
\caption{Schematics of hcp lattice (Euler angles $\phi_1=\SI{0}{\degree}, \Phi=\SI{0}{\degree}, \phi_2=\SI{0}{\degree}$) transforming to bcc lattice (one scenario, $\theta = \pm\SI{5.26}{\degree}$)}
\label{fig: schematics hcp-bcc transformation}
\end{figure}

The transformation from close-packed hexagonal (hcp) to body-centered cubic (bcc) crystal structure is accommodated by modulating atoms on the basal plane to reach the stacking of bcc \hkl{110} planes such that the original basal plane transforms to one of the \hkl{110} planes in bcc structure.
According to hexagonal symmetry, the transformed \hkl{110} plane has 3 possible orientations.
For the ideal \(c/a\) ratio, the height of hexagonal unit cells remains constant but the intermediate atoms and some atoms on the original basal plane slightly rearrange their positions to become the corner atoms of the bcc unit cells.
This process can be decomposed to a \SI{5.26}{\degree} rotation (either clockwise or counterclockwise) $\tnsr R_a$/$\tnsr R_b$ ($a,b$ means the rotation in the 1st and 2nd step, respectively) around the basal plane normal, \ie, \crystaxis{c} direction, to align the long diagonal of the hexagon (\hkl<2-1-10> direction) with bcc \hkl<111> direction, and a stretch of hcp \hkl<1-210> and a contraction along \hkl<10-10>, see \cref{fig: schematics hcp-bcc transformation}.

The transformation from bcc to hcp is the reversal of the above process, whereas there are 6 equivalent \hkl{110} planes to choose to form the new basal plane in the hcp unit cell according to bcc symmetries.
Thus, the parent $\beta$ and its transformed $\alpha$ variants (vice versa) preserve specific orientation relationships called Burgers Orientation Relationship (BOR) \citep{Burgers1934}.

The method to calculate lattice orientation in hcp-bcc-hcp phase transformation process is summarized as below.%
\footnote{Since $\alpha$ phase typically dominates the microstructure at ambient temperature, comparisons among different $\alpha$ grains undergone hcp-bcc-hcp phase transformation are of great interest.}
In addition, a statistical analysis of variants associated with this transformation is also presented in this section.

%\question{what about non-ideal \(c/a\) ratios}

\subsection{Deformation gradient calculation}
\label{sec: F calculation}
The deformation gradient calculations are essentially doing two shape changes in the correct frames.
The two steps are summarized as follows:
\begin{itemize}
\item hcp$\to$bcc: the first shape change $\transpose{\tnsr R_a}\,\tnsr U$ is composed of an active plane rotation $\transpose{\tnsr R_a}$ and an active plane stretch \tnsr U to transform from the hcp basal plane to the bcc \hkl{110} plane in the standard hexagonal frame $\tnsr R_\text{fr1}$.
\begin{align} 
\tnsr F_1 &:=\: 
\underbrace{
\transpose{\ _\text{lab}^\text{hex}\tnsr{G}}
\transpose{\tnsr{S}_\text{hex}}
}_{\transpose{\tnsr R_\text{fr1}}}\ 
\transpose{\tnsr R_a} \tnsr{U}\
\underbrace{   
\tnsr{S}_\text{hex}
\ _\text{lab}^\text{hex}\tnsr{G} 
}_{\tnsr R_\text{fr1}}
\end{align}

\item bcc$\to$hcp: the second shape change $\transpose{\tnsr R_b}\ \inverse{\tnsr U}$ includes an active plane rotation $\transpose{\tnsr R_b}$ and an operation to undo the stretch such that it transforms the bcc \hkl{110} plane back to the hcp basal plane shape in the transformed bcc frame $\tnsr R_\text{fr2}$.
\begin{align}
\begin{split}
\tnsr F_2 :=\: 
\underbrace{
\transpose{_\text{lab}^\text{hex}\tnsr{G}} 
\transpose{\tnsr{S}_\text{hex}}\  
\transpose{\tnsr R_a}\  
\transpose{_\text{110}^\text{bcc}\tnsr{R}}\  
\transpose{\tnsr{S}_\text{bcc}}\  
\transpose{_\text{bcc}^{110}\tnsr{R}} 
}_{\transpose{\tnsr R_\text{fr2}}}\ 
\transpose{\tnsr R_b} 
\inverse{\tnsr{U}}  \\
\underbrace{
\ _\text{bcc}^\text{110}\tnsr{R}\ 
\tnsr{S}_\text{bcc}  
\ _\text{110}^\text{bcc}\tnsr{R}\ 
 \tnsr R_a\  
\tnsr{S}_\text{hex}\ 
\ _\text{lab}^\text{hex}\tnsr{G}\  
}_{\tnsr R_\text{fr2}}
\end{split}
\end{align}

\end{itemize}

The combined calculations of the deformation gradient  $\tnsr{F}$ are as follows. 
\begin{align}
\tnsr{F} &:=\: \tnsr F_2\ \tnsr F_1
%\\
%&:=\: \transpose{_\text{lab}^\text{hex}\tnsr{G}}
%\transpose{\tnsr{S}_\text{hex}} 
%\underbrace{
%\transpose{\tnsr R_a} 
%\transpose{_\text{110}^\text{bcc}\tnsr{R}} 
%\transpose{\tnsr{S}_\text{bcc}} 
%\transpose{_\text{bcc}^{110}\tnsr{R}}
% }_{\transpose{\tnsr M}} 
%\transpose{\tnsr R_b} 
%\inverse{\tnsr{U}}
%\underbrace{
%\ _\text{bcc}^\text{110}\tnsr{R}\ 
%\tnsr{S}_\text{bcc} 
%\ _\text{110}^\text{bcc}\tnsr{R}\ 
% \tnsr R_a 
% }_{\tnsr M}
%\overbrace{  
%\tnsr{S}_\text{hex}
%\ _\text{lab}^\text{hex}\tnsr{G} 
%\transpose{\ _\text{lab}^\text{hex}\tnsr{G}}
%\transpose{\tnsr{S}_\text{hex}}
%}^{\equiv\,\I} 
%\transpose{\tnsr R_a} \tnsr{U}\   
%\tnsr{S}_\text{hex}
%\ _\text{lab}^\text{hex}\tnsr{G} 
\end{align}

$\textbf{F}$ varies with a different initial orientation and by applying different symmetry operators.
Note the rotation matrix $\tnsr R_a/\tnsr R_b$ and the shape change operator $\tnsr U$ in \cref{fig: schematics hcp-bcc transformation} are determined based on the hcp crystal.
For alloys with varying alloying compositions, a specific $c/a$ ratio should be applied to ensure more accurate orientation and deformation gradient calculations. 

\Cref{tab:57variants} in \cref{sec: appendix} summarizes the orientations and shape changes (leftmost cubes) of these 57 variants.
The top three rows show the transformations with the unit symmetry operation in both steps, resulting in an invariant case and essentially pure rotations in two different directions

\subsection{Orientation calculation}
The orientation calculation in hcp-bcc-hcp transformation follows a series of rotational and frame transformations, which was first introduced by \citet{Humbert_etal1995}.

\begin{enumerate}
	\item Convert original Euler angles\footnote{Quaternions or any other conventions to represent crystal orientation in the lab frame can also be used.} to the orientation matrix (lab to crystal frame), $_\text{lab}^\text{hex}\tnsr G$.
	\item Apply hexagonal symmetry operators (see \cref{tab:symhex}) in the hexagonal frame $\tnsr{S}_\text{hex}$.
	\item Apply rotation $\tnsr R_a$ around \crystaxis{c} (clockwise ``$-$'' and counter clockwise ``$+$'') to align the hcp \hkl<2-1-10> direction of the hcp basal plane with the corresponding bcc \hkl<111> direction (bcc \hkl{110} plane diagonal), which essentially transforms the hexagonal frame to the bcc \hkl{110} plane frame. 
	\item Convert from the bcc \hkl{110} plane frame to the default bcc frame, $_{110}^\text{bcc}\tnsr R$.%
	\footnote{Non-ideal \(c/a\) ratio would result in a different \(_{110}^\text{bcc}\tnsr R\). 
	%\question{Will the transformed bcc be bct for non-ideal ratio} yes it will but will it affect the symmetry operators too?}
	}
	\item Apply cubic symmetry operators $\tnsr{S}_\text{bcc}$  (see \cref{tab:symbcc}) in the bcc frame.
	\item Convert from the bcc frame back to the bcc \hkl{110} plane frame, $\transpose{_{110}^\text{bcc}\tnsr R}$ (equivalent to $_\text{bcc}^{110}\tnsr R$).
	\item To start the bcc-hcp transformation, apply in-plane rotation $\tnsr R_b$ to align the bcc \hkl<111> direction with the hcp \hkl<2-1-10> direction, which essentially transforms the bcc \hkl{110} plane frame back to the hexagonal frame.
	The new orientation matrix $\overset{\text{new}}{\tnsr G}$ expressed in the lab coordinates %\question{why ``expressed in lab'' when it translates lab to crystal} after the hcp-bcc-hcp transformation is calculated as follows.
	\begin{align*}
		\overset{\text{new}}{\tnsr{G}} &:=\: \transpose{_\text{lab}^\text{hex}\tnsr G}\ \tnsr R_b \ _{110}^\text{bcc}\tnsr R^\text{T}\ \tnsr S_\text{bcc}\ _{110}^\text{bcc}\tnsr R\ \tnsr R_a\ \tnsr S_\text{hex}\ _\text{lab}^\text{hex}\tnsr G
	\end{align*}
\end{enumerate}

\section{Automated bcc reconstruction}
\label{sec: automated reconstruction}
%============================================================================
\subsection{Reconstruction algorithm}

\begin{figure*}
\centering
\includegraphics[width=.9\linewidth]{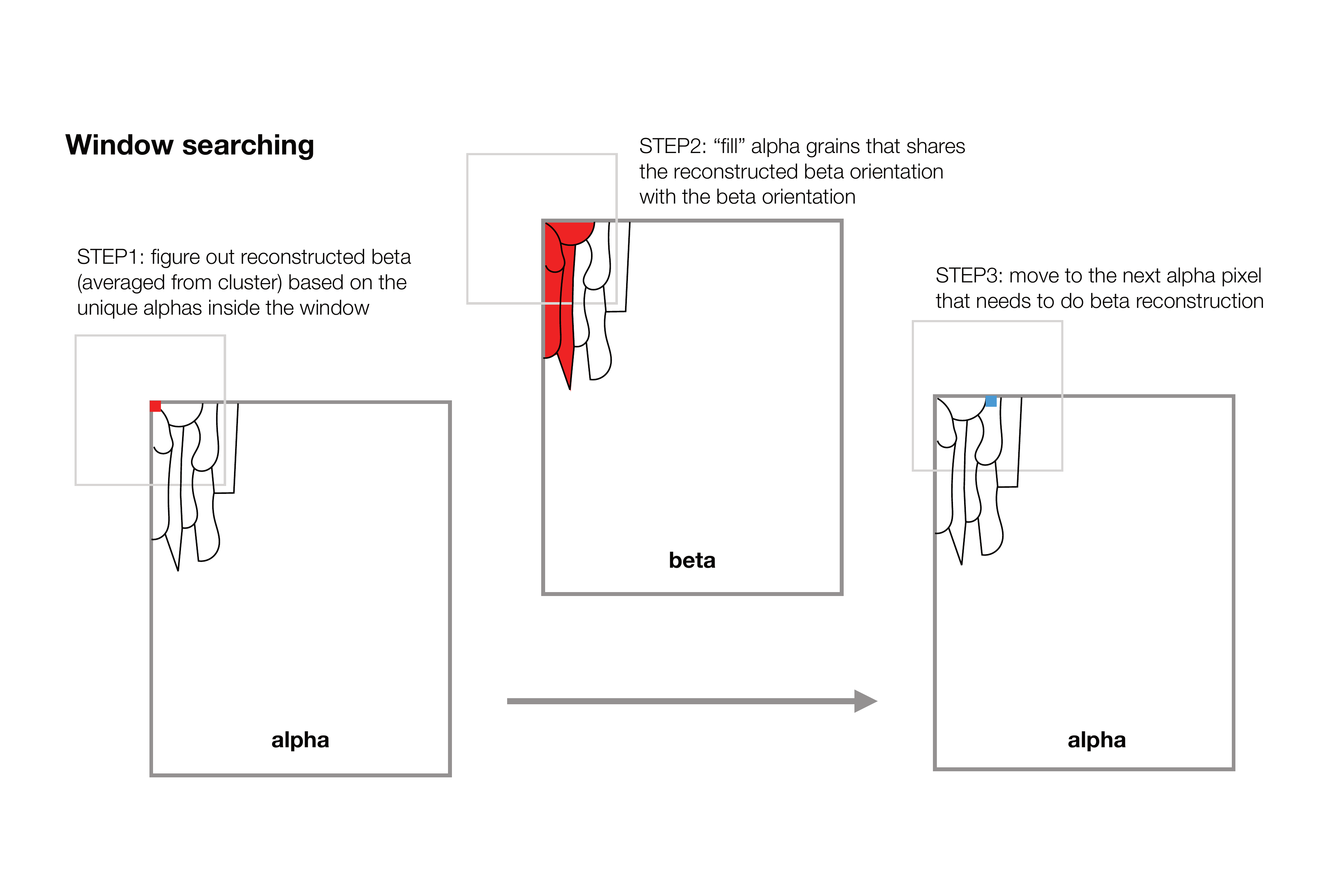}
\caption{Schematic of reconstructing ``parent'' beta grains by a window searching algorithm.}
\label{fig: beta recon algorithm}
\end{figure*}

This section presents a $\beta$ cluster-averaging approach to facilitate automated $\beta$ reconstruction from EBSD mapping in \cref{sec: automated reconstruction} and validation of the approach based on synthetic and measured data in \cref{sec: validation_sim,sec: validation_exp}.
Since the new approach works by grouping $\beta$ orientations, it should be an accurate method. 
On the other hand, this new approach uses a window searching algorithm to pre-select spatially-close $\alpha$ grains (candidates) likely coming from the same parent grain.
The window searching algorithm narrows down the $\alpha$ population per reconstruction unit and inherently supports parallel execution, which can greatly accelerate the $\beta$ reconstruction of the whole map.
Therefore, the new cluster-averaging approach has a good balance of accuracy and efficiency.
The algorithm goes as follows:

\begin{figure*}
\centering
\includegraphics[width = \linewidth]{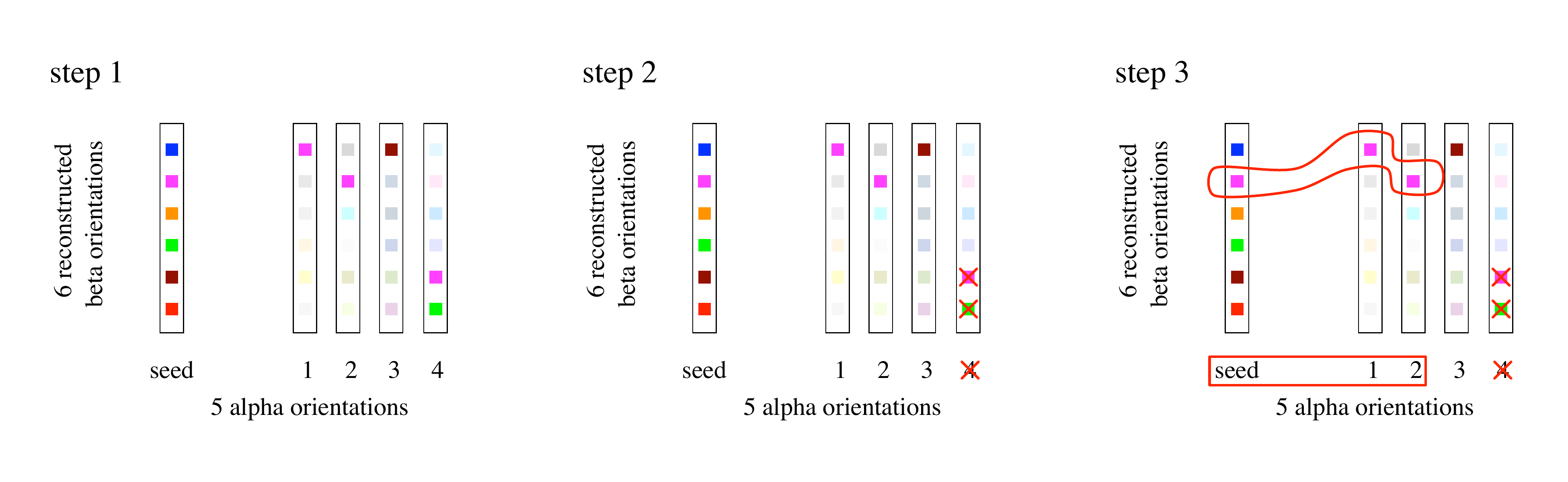}
\caption{The first step: dynamically adding $\beta$ orientations to the cluster by checking each $\beta$ orientation of the ``seed'' against the $\beta$ orientation of the other $\alpha$ orientations (columns). The second step excludes the members violating the unique determination rule. The third step finds the longest shared $\beta$ cluster (magenta).}
\label{fig: checkInStack}
\end{figure*}

\begin{figure}[h!]
\centering
\includegraphics[width=.3\linewidth]{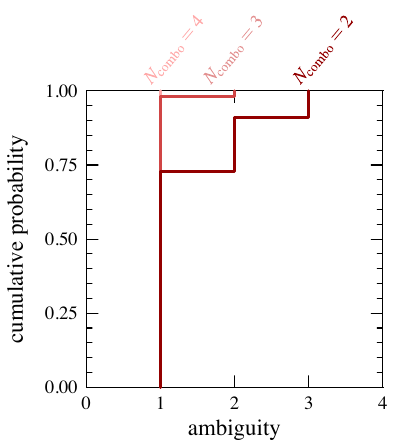}
\caption{The ``parent'' beta orientation is uniquely determined when  $\N{combo} \geq 4$.}
\label{fig: ambiguity}
\end{figure}

\begin{enumerate}
\item Grains (grainID) of the whole map are identified with Floodfill based on misorientation angle within \SI{5}{\degree}. This results in arbitrarily many (N) grainIDs (with associated average orientations).

\item Cluster N grainIDs into \(\text{n} <= \text{N}\) distinct orientationIDs (using again misorientation threshold \SI{5}{\degree}) maintaining a mapping i elem N to j elem n (with associated average orientations for each orientationID).

\item Back-calculate \(\text{p} \times \text{n}\) potential parent orientations (\(\text{p}=6\) for \(\beta\) reconstruction).

\item Select the top-left most pixel for which the grainID has not had the parent orientation identified yet.

\item Select m by m pixel window centered on the pixel from 4, see \cref{fig: beta recon algorithm}.

\item For each of that grainID's p possible parent orientations (based on its orientationID), count how many other grains also (and exclusively) share (within an orientation tolerance \SI{5}{\degree}) that potential parent, \ie, the intersection of parent orientations between any of the identified ``partners'' and the central ``seed'' orientationID needs to comprise only the currently selected orientation. 
For example, alpha \#4 (column 4 in \cref{fig: checkInStack} middle) has two identified $\beta$ orientations (magenta and green $\beta$ orientation), which is not a uniquely determined situation.
This process is repeated with updated average group parent orientation until no further new members are found.

\item Identify the largest group with len \(>1\) of shared parent orientation.%
\footnote{To understand how many $\alpha$ variants are needed to determine a unique $\beta$ orientation and whether the determined $\beta$ orientation is correct: a synthetic $\beta$ is used to produce 12 $\alpha$ orientations; then the possible parent $\beta$ orientations shared by a group ($\N{combo}=2,3,4,..,12$) of these 12 $\alpha$ orientations are calculated and compared to the original $\beta$ orientation. 
\Cref{fig: ambiguity} shows the number of possible parent $\beta$ orientations determined from varying $\alpha$ group sizes, and it is found that the parent $\beta$ orientation is uniquely determined when  $\N{combo} \geq 4$. 
It is also found that the uniquely determined $\beta$ orientation is always correct.
Thus, a strategy to do $\beta$ reconstruction is finding the unique $\beta$ orientation shared by the largest $\alpha$ group size, called the unique determination rule. }
\Cref{fig: checkInStack} right shows that the magenta $\beta$ cloud is the ``longest'' shared cluster (shared by the seed, alpha \#1, and alpha \#2).

\item Store this parent orientation (variant of own orientationID) for every grainID in the longest group, see step 2 in \cref{fig: beta recon algorithm}.

\item Goto 4.
\end{enumerate}

%===========================================================================================
\subsection{Validation of bcc reconstruction based on synthetic data}
\label{sec: validation_sim}
%===========================================================================================

\begin{figure}[h!]
\centering
\minipage[t]{0.2\textwidth}
	\centering
	\includegraphics[width=.8\linewidth,valign=t]{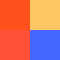}
  	\caption*{parent $\beta$ orientations}
\endminipage
\minipage[t]{0.2\textwidth}
	\centering
  	\includegraphics[width=.8\linewidth,valign=t]{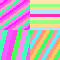}
 	\caption*{children $\alpha$ orientations }
\endminipage
\minipage[t]{0.2\textwidth}
	\centering
	\includegraphics[width=.8\linewidth,valign=t]{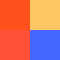}
	\caption*{reconstructed $\beta$ orientations}
\endminipage
\caption{Correct \(\beta\) reconstruction results based upon synthetic data.}
\label{fig: synthetic}
\end{figure}

The above implementation was first tested with synthetic data made with four blocks of random parent $\beta$ orientations for each, as shown in \cref{fig: synthetic} left. 
Stripe-shape% 
\footnote{Note that the reconstruction is based on the distinct hexagonal orientations found within the window. The grain geometry, however, does not directly influence the results. }
child $\alpha$ grains (\cref{fig: synthetic} middle) were then generated to mimic the secondary alpha grains commonly observed in many Ti alloys, such as Ti-6Al-2Sn-4Zr-6Mo (w.t.\%) as a result of variant selections. 
\Cref{fig: synthetic} right shows the reconstructed orientations for $\beta$ grains. 
They are consistent with the \emph{actual} parent $\beta$ orientations in \cref{fig: synthetic} left.

%===========================================================================================
\subsection{Validation of bcc reconstruction based on measured data}
\label{sec: validation_exp}
%===========================================================================================
\begin{figure}
\centering
\minipage{0.24\textwidth}
	\centering
	\includegraphics[valign=t, width=.9\linewidth]{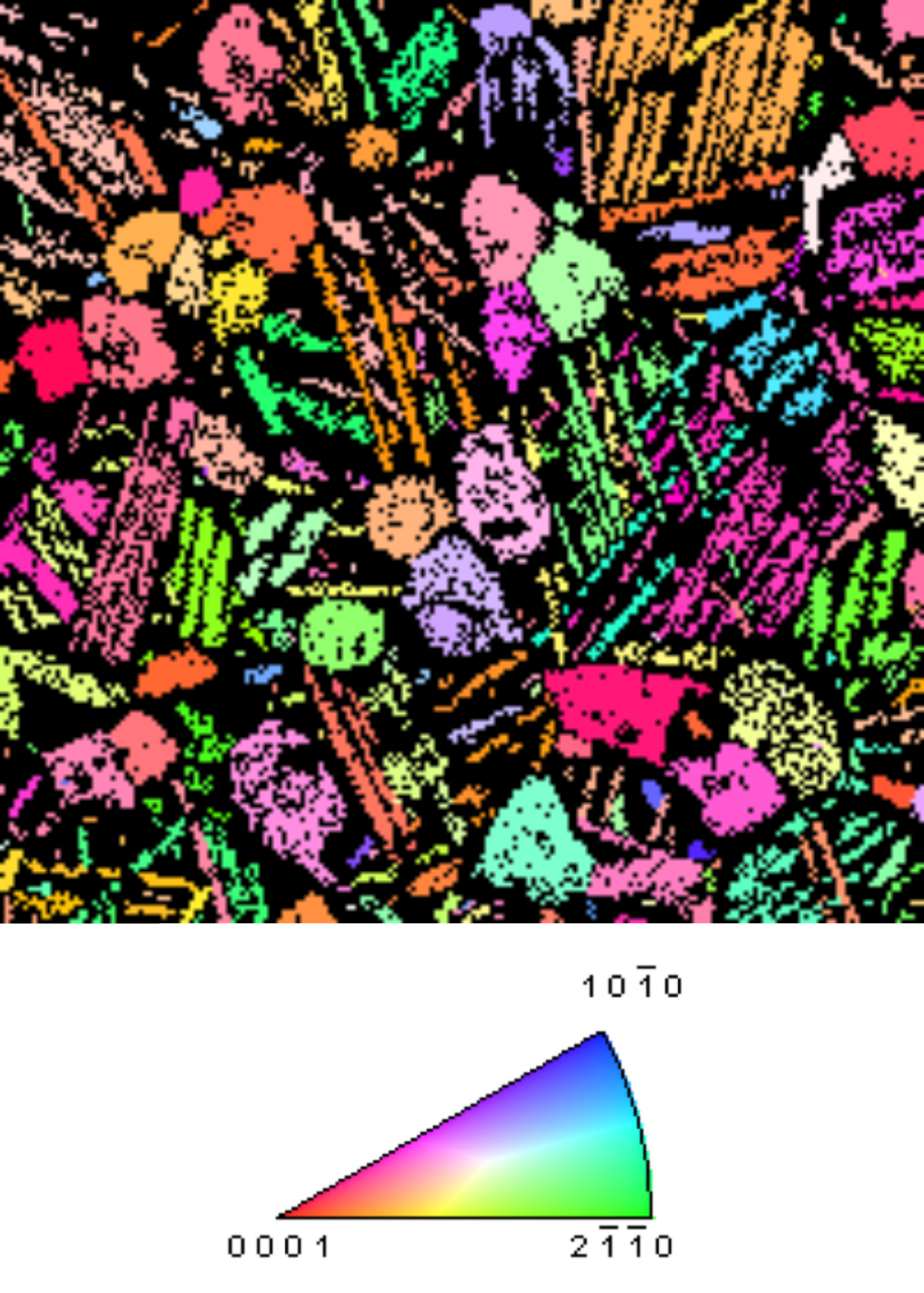}
  	\caption*{measured $\alpha$ IPF001}
\endminipage
\minipage{0.24\textwidth}
	\centering
	\includegraphics[valign=t,width=.9\linewidth]{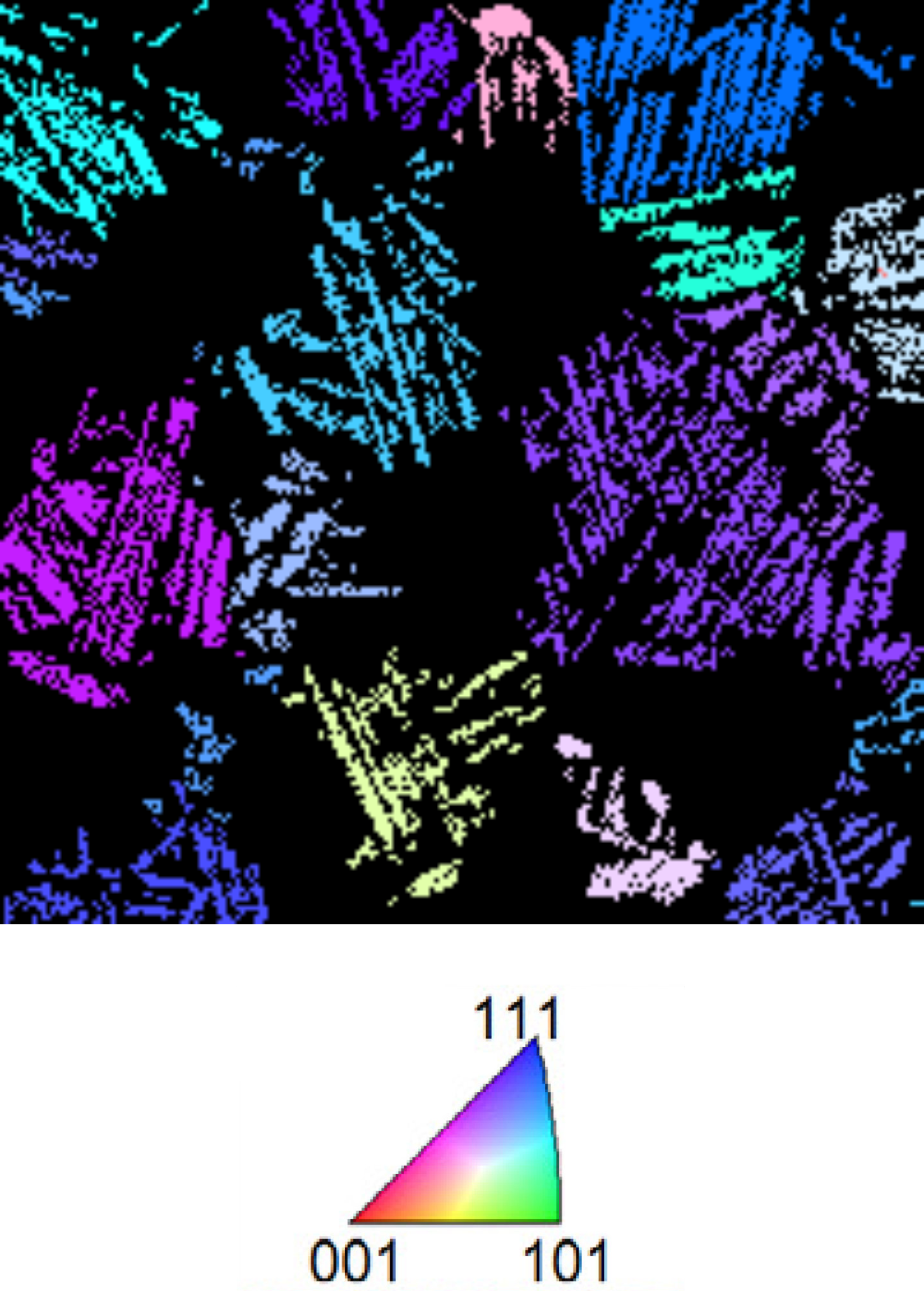}
	\caption*{reconstructed $\beta$ of secondary $\alpha$ IPF001 }
\endminipage
\caption{$\beta$ reconstruction (200 x 200) of lamella $\alpha$ phase in Ti-6Al-2Sn-4Zr-6Mo using window size of 71 (original data provided by \citep{Banerjee2015}). }
\label{fig: Ti6246}
\end{figure}

A measured orientation map of the $\alpha$ phase in a \Tisixtwo\ (w.t.\%) sample measured by electron backscattered diffraction (EBSD) was used to test the robustness of the automated $\beta$ reconstruction implementation. 
The sample has near-equiaxed primary alpha grains and plate-like secondary $\alpha$ grains that form a \emph{typical} ``basket weave'' microstructure resulting from variant selections in the $\beta\to\alpha$ transformation (left in \cref{fig: Ti6246}). 

The reconstructed $\beta$ orientations for lamella $\alpha$ grains are presented in \cref{fig: Ti6246} right.
The interwoven grains from the same $\beta$ parent grain show the same orientation color in the reconstruction results.

%=========================================
\section{Summary}
\label{sum5}
%=========================================
This study proposed a cluster-averaging approach to automate the reconstruction of $\beta$ orientations from $\alpha$ orientations. 
The approach was successfully applied to reconstruct the $\beta$ orientation map from the lamellar microstructure of the $\alpha$ phase in a Ti-6Al-2Sn-4Zr-6Mo sample, demonstrating its effectiveness for complex microstructural analysis during hcp-bcc-hcp phase transformation.
Additionally, a method was introduced to calculate the shear deformation (deformation gradient \vctr F) during hcp-bcc phase transformation. 
This method enables the seamless integration of transformation kinematics into continuum-based models by providing a computationally efficient approach to determine the deformation gradient influencing the transformation.

\section*{Acknowlegement}
Funding from U.S. National Science Foundation DMR-1411102 is greatly appreciated.
The authors would like to thank Dr.~Dipankar Banerjee and Dr.~Arunima Banerjee from Indian Institute of Science for providing the measured orientation data.

\section{Code Availability}
Python scripts of both the deformation gradient calculation and cluster-averaging approach can be found at \url{https://github.com/zhuowenzhao/BetaReconstruction}.

\bibliographystyle{elsarticle-harv}
\bibliography{Zhao_etal2022}

\section*{Appendix}
\label{sec: appendix}

\begin{table}[h!]
\caption{List of 12 hexagonal lattice symmetry operators used in the thesis ($a = \frac{\sqrt{3}}{2} $)}
\label{tab:symhex}
\begin{center}
%\begin{tabular}{ M{4cm} M{4cm} M{4cm} } 
\begin{tabular}{ cccccc } 
\toprule
 $\text{h}_{1}$ & 
 $\begin{bmatrix} 1&0&0 \\ 0&1&0 \\ 0&0&1 \end{bmatrix}$ &
 $\text{h}_{2}$ & 
 $\begin{bmatrix} -0.5&a&0 \\ -a&-0.5& 0 \\ 0&0&1 \end{bmatrix}$ &
 $\text{h}_{3}$ &
 $\begin{bmatrix} -0.5&-a&0 \\ a&-0.5&0 \\ 0&0&1 \end{bmatrix}$\\[1cm] 
 
 $\text{h}_{4}$ & 
 $\begin{bmatrix} 0.5&a&0 \\ -a&0.5&0 \\ 0&0&1 \end{bmatrix}$ &
 $\text{h}_{5}$ & 
 $\begin{bmatrix} -1&0&0 \\ 0&-1& 0 \\ 0&0&1 \end{bmatrix}$ &
 $\text{h}_{6}$ &
 $\begin{bmatrix} 0.5&-a&0 \\ a&0.5&0 \\ 0&0&1 \end{bmatrix}$\\[1cm] 
 
 $\text{h}_{7}$ & 
 $\begin{bmatrix} -0.5&-a&0 \\ -a&0.5&0 \\ 0&0&1 \end{bmatrix}$ &
 $\text{h}_{8}$ & 
 $\begin{bmatrix} -1&0&0 \\ 0&-1& 0 \\ 0&0&1 \end{bmatrix}$ &
 $\text{h}_{9}$ &
 $\begin{bmatrix} 0.5&-a&0 \\ a&0.5&0 \\ 0&0&1 \end{bmatrix}$\\[1cm] 
 
 $\text{h}_{10}$ & 
 $\begin{bmatrix} 0.5&a&0 \\ a&-0.5&0 \\ 0&0&-1 \end{bmatrix}$ &
 $\text{h}_{11}$ & 
 $\begin{bmatrix} -1&0&0 \\ 0&1& 0 \\ 0&0&-1 \end{bmatrix}$ &
 $\text{h}_{12}$ &
 $\begin{bmatrix} 0.5&-a&0 \\ -a&-0.5&0 \\ 0&0&-1 \end{bmatrix}$\\[1cm] 

\bottomrule
\end{tabular}
\end{center}
\end{table}

\begin{center}
\captionof{table}{List of 57 possible variants in hcp-bcc-hcp phase transformation. $\text{b}_i$ is the bcc symmetry operator in \cref{tab:symbcc}, and $\text{h}_i$ is the hexagonal symmetry operator in \cref{tab:symhex}. ``$+$'' and ``$-$'' denote clockwise and counterclockwise rotation, respectively. The cubes in the rightmost column show the transformed cubic volume (the original shape is the same as the first cube). \label{tab:57variants}}
\tablehead{%
    \toprule 
   hex & $R_a$& bcc & $R_b$ & $\Delta\theta$ (\SI{}{\degree}) & new orientation & shape \\ 
    \hline}
\tabletail{\hline}

\xentrystretch{0.03}
\setlength\extrarowheight{1pt} % for a more open "look"
   
\begin{xtabular}{*{7}{c}}                 
 & $+$ & & $-$ &\num{0.0} & \includegraphics[scale=0.1,align=c]{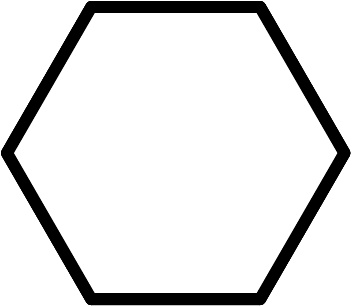} & \includegraphics[scale=0.1,align=c]{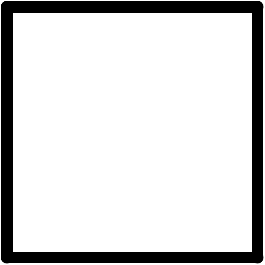} \\[5pt] 
\multirow{2}*{$\text{h}_{1}$} & $+$ & \multirow{2}*{ $\text{b}_{1}$ } & $+$ &\num{10.52877936550927} & \includegraphics[scale=0.1,align=c]{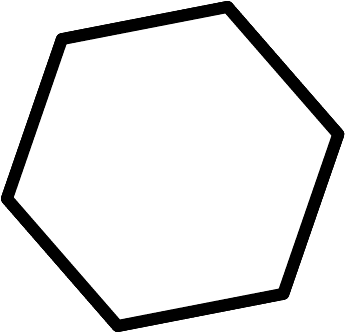} & \includegraphics[scale=0.1,align=c]{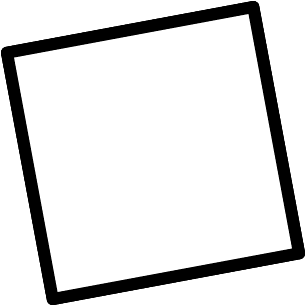} \\[5pt] 
 & $-$ & & $-$ &\num{10.52877936550927} & \includegraphics[scale=0.1,align=c]{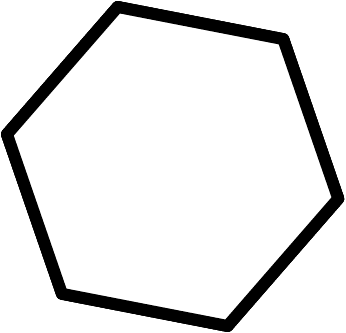} & \includegraphics[scale=0.1,align=c]{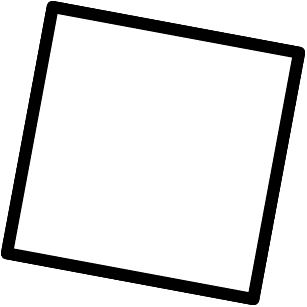} \\[5pt] 

 \hline 
 & $+$ & & $-$ &\num{90.00000000000001} & \includegraphics[scale=0.1,align=c]{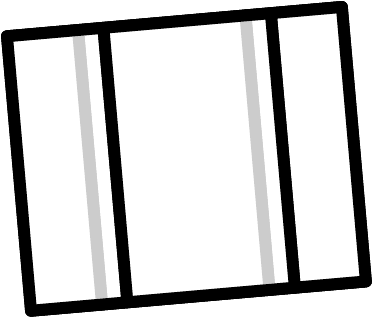} & \includegraphics[scale=0.1,align=c]{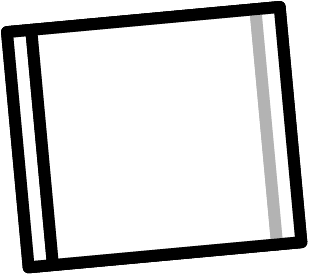} \\[5pt] 
\multirow{2}*{$\text{h}_{1}$} & $+$ & \multirow{2}*{ $\text{b}_{2}$ } & $+$ &\num{90.00000000000001} & \includegraphics[scale=0.1,align=c]{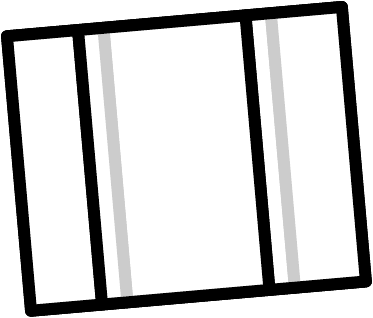} & \includegraphics[scale=0.1,align=c]{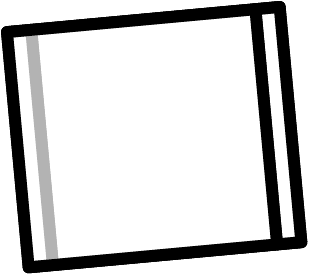} \\[5pt] 
 & $-$ & & $-$ &\num{90.00000000000001} & \includegraphics[scale=0.1,align=c]{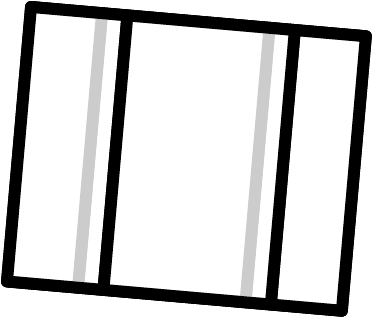} & \includegraphics[scale=0.1,align=c]{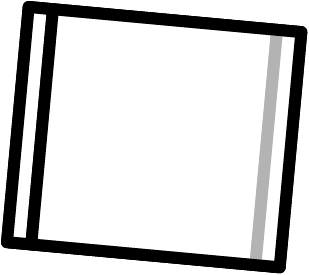} \\[5pt] 
 & $-$ & & $+$ &\num{90.00000000000001} & \includegraphics[scale=0.1,align=c]{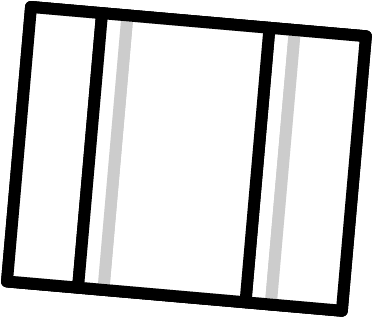} & \includegraphics[scale=0.1,align=c]{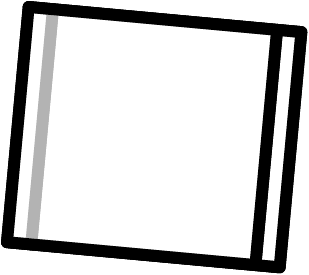} \\[5pt] 
 
 \hline 
 & $+$ & & $-$ &\num{60.000000000000036} & \includegraphics[scale=0.1,align=c]{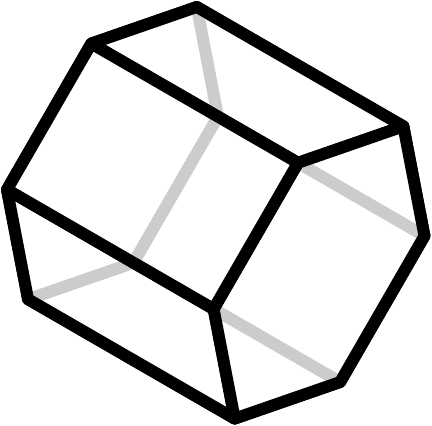} & \includegraphics[scale=0.1,align=c]{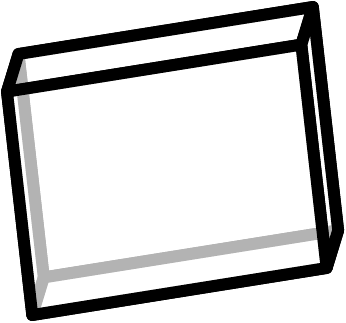} \\[5pt] 
\multirow{2}*{$\text{h}_{1}$} & $+$ & \multirow{2}*{ $\text{b}_{3}$ } & $+$ &\num{60.83197478497545} & \includegraphics[scale=0.1,align=c]{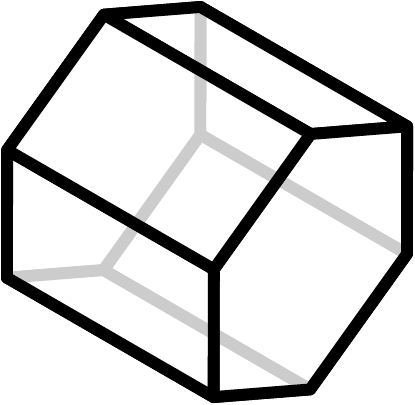} & \includegraphics[scale=0.1,align=c]{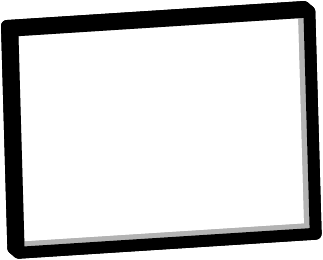} \\[5pt] 
 & $-$ & & $-$ &\num{60.83197478497545} & \includegraphics[scale=0.1,align=c]{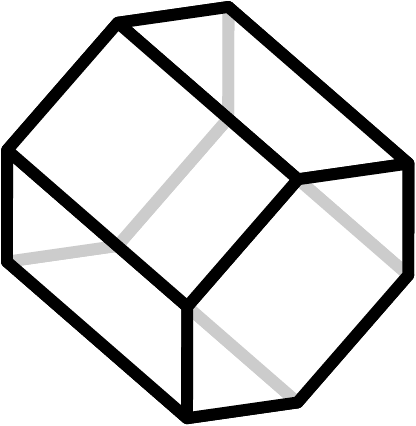} & \includegraphics[scale=0.1,align=c]{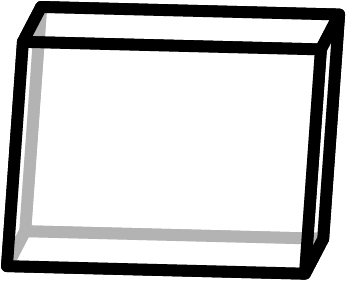} \\[5pt] 
 & $-$ & & $+$ &\num{63.26177218039435} & \includegraphics[scale=0.1,align=c]{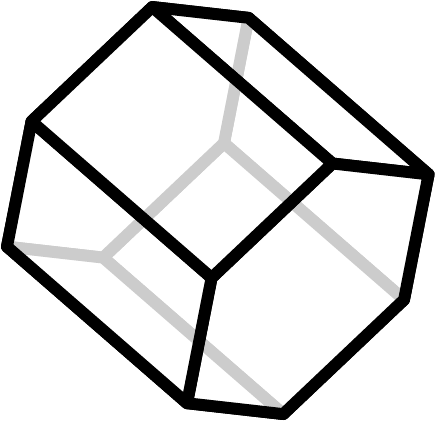} & \includegraphics[scale=0.1,align=c]{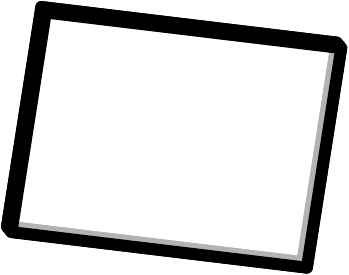} \\[5pt] 

 \hline 
 & $+$ & & $-$ &\num{63.26177218039435} & \includegraphics[scale=0.1,align=c]{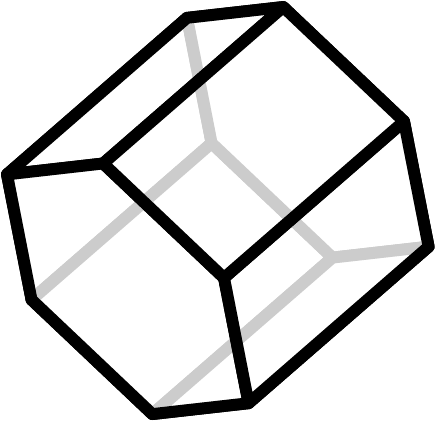} & \includegraphics[scale=0.1,align=c]{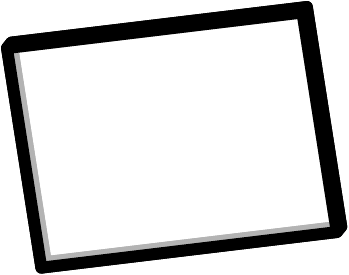} \\[5pt] 
\multirow{2}*{$\text{h}_{1}$} & $+$ & \multirow{2}*{ $\text{b}_{4}$ } & $+$ &\num{60.83197478497545} & \includegraphics[scale=0.1,align=c]{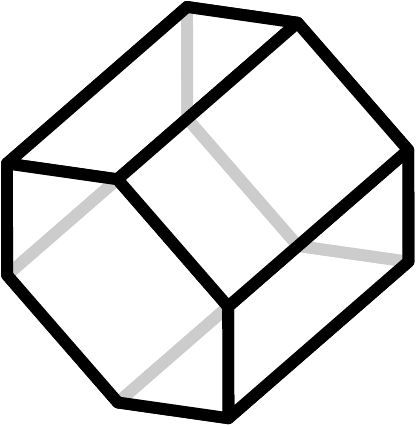} & \includegraphics[scale=0.1,align=c]{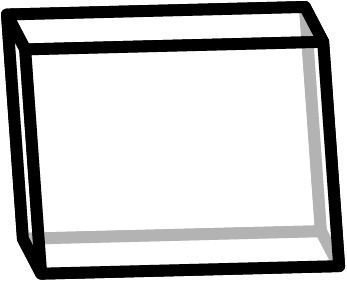} \\[5pt] 
 & $-$ & & $-$ &\num{60.83197478497545} & \includegraphics[scale=0.1,align=c]{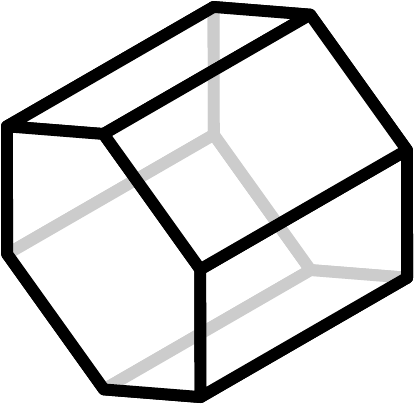} & \includegraphics[scale=0.1,align=c]{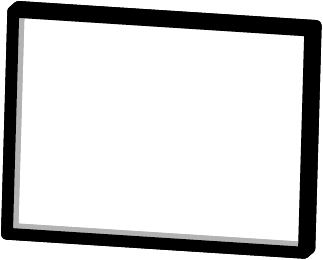} \\[5pt] 
 & $-$ & & $+$ &\num{60.000000000000036} & \includegraphics[scale=0.1,align=c]{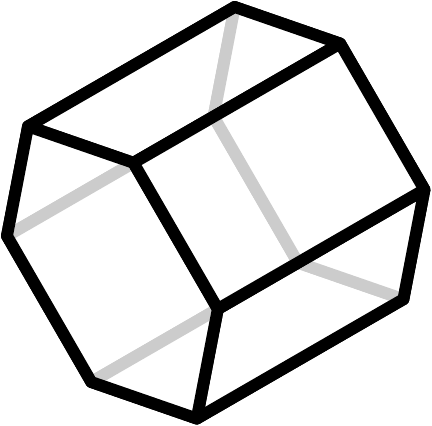} & \includegraphics[scale=0.1,align=c]{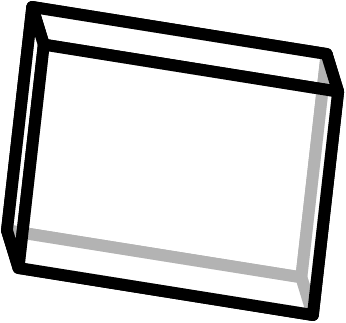} \\[5pt] 

 \hline 
 & $+$ & & $-$ &\num{60.000000000000036} & \includegraphics[scale=0.1,align=c]{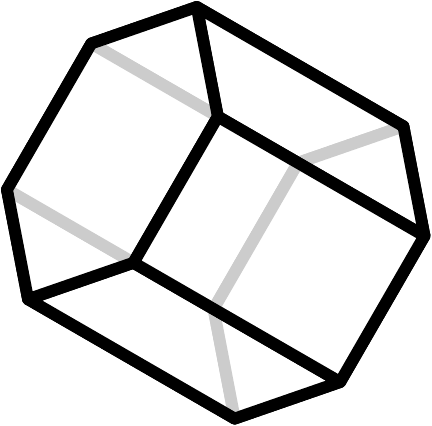} & \includegraphics[scale=0.1,align=c]{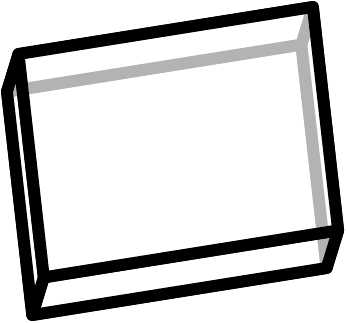} \\[5pt] 
\multirow{2}*{$\text{h}_{1}$} & $+$ & \multirow{2}*{ $\text{b}_{9}$ } & $+$ &\num{60.83197478497545} & \includegraphics[scale=0.1,align=c]{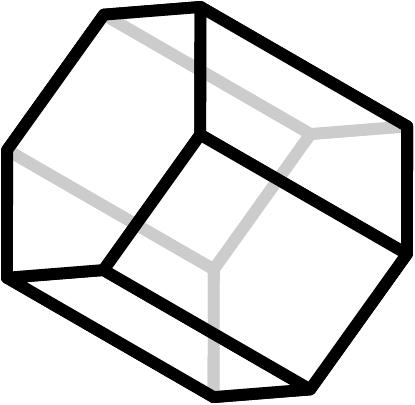} & \includegraphics[scale=0.1,align=c]{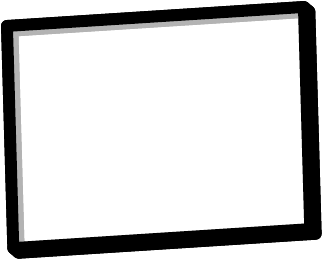} \\[5pt] 
 & $-$ & & $-$ &\num{60.83197478497545} & \includegraphics[scale=0.1,align=c]{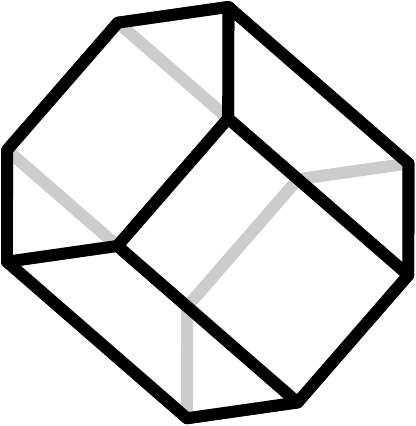} & \includegraphics[scale=0.1,align=c]{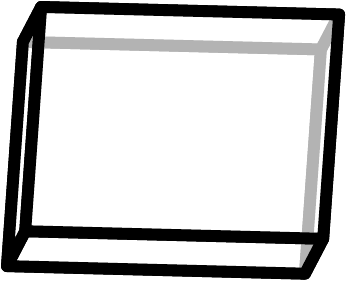} \\[5pt] 
 & $-$ & & $+$ &\num{63.2617721803944} & \includegraphics[scale=0.1,align=c]{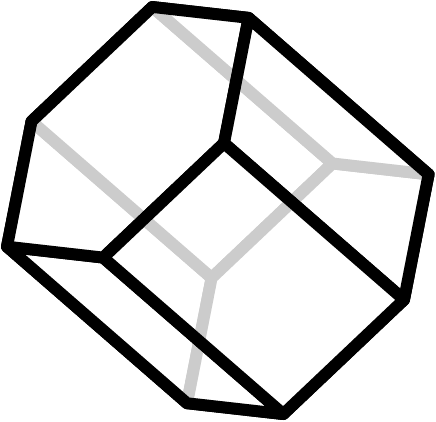} & \includegraphics[scale=0.1,align=c]{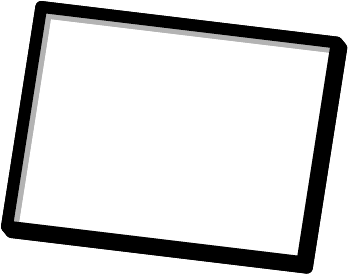} \\[5pt] 

 \hline 
 & $+$ & & $-$ &\num{63.2617721803944} & \includegraphics[scale=0.1,align=c]{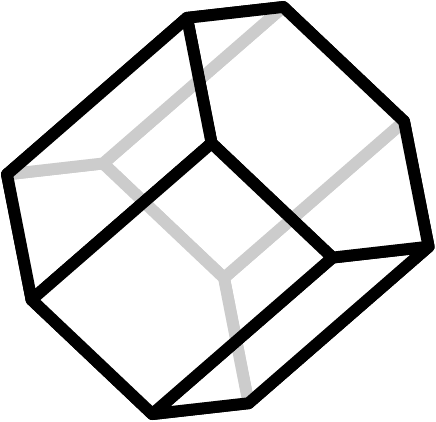} & \includegraphics[scale=0.1,align=c]{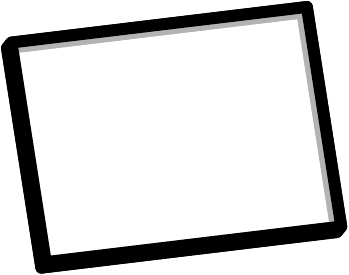} \\[5pt] 
\multirow{2}*{$\text{h}_{1}$} & $+$ & \multirow{2}*{ $\text{b}_{10}$ } & $+$ &\num{60.83197478497545} & \includegraphics[scale=0.1,align=c]{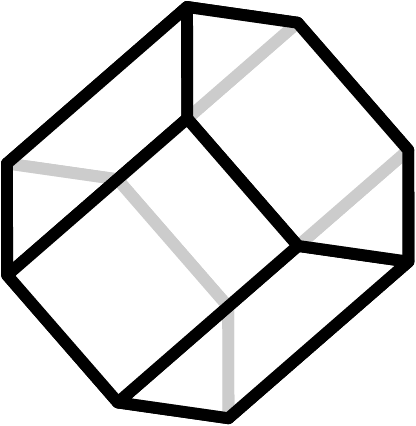} & \includegraphics[scale=0.1,align=c]{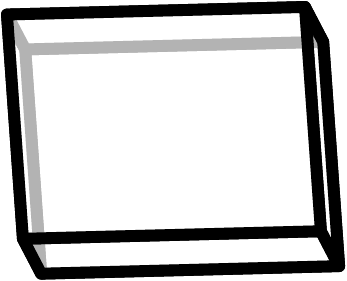} \\[5pt] 
 & $-$ & & $-$ &\num{60.83197478497545} & \includegraphics[scale=0.1,align=c]{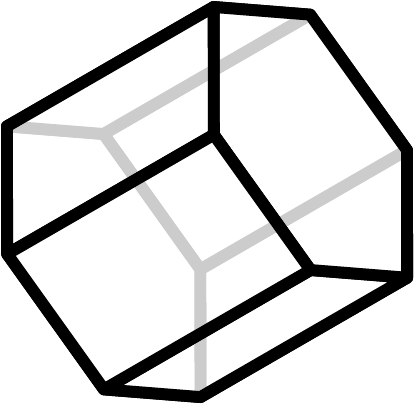} & \includegraphics[scale=0.1,align=c]{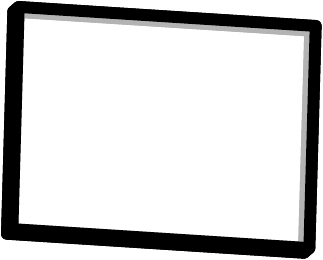} \\[5pt] 
 & $-$ & & $+$ &\num{60.000000000000036} & \includegraphics[scale=0.1,align=c]{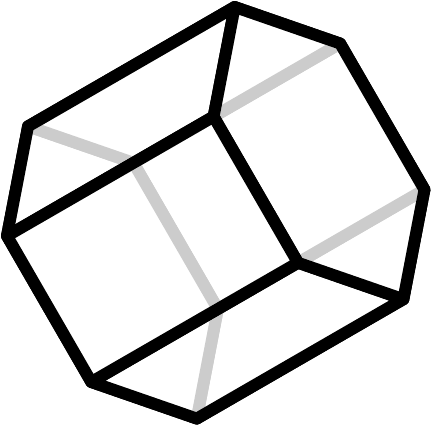} & \includegraphics[scale=0.1,align=c]{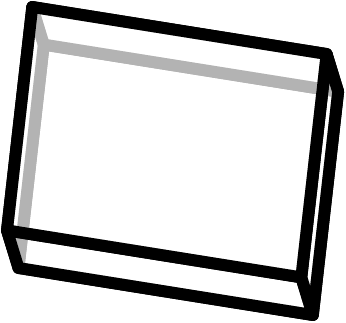} \\[5pt] 

 \hline 
 & $+$ & & $-$ &\num{90.00000000000001} & \includegraphics[scale=0.1,align=c]{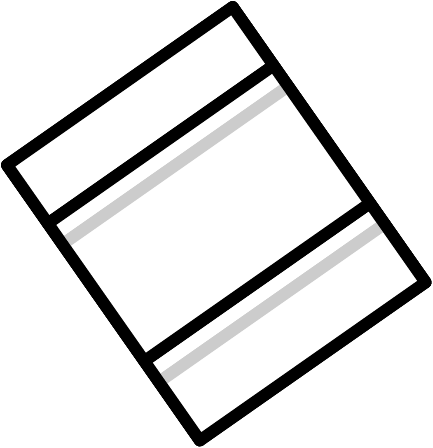} & \includegraphics[scale=0.1,align=c]{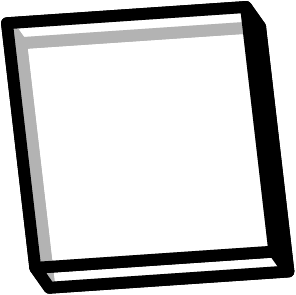} \\[5pt] 
\multirow{2}*{$\text{h}_{2}$} & $+$ & \multirow{2}*{ $\text{b}_{2}$ } & $+$ &\num{90.00000000000001} & \includegraphics[scale=0.1,align=c]{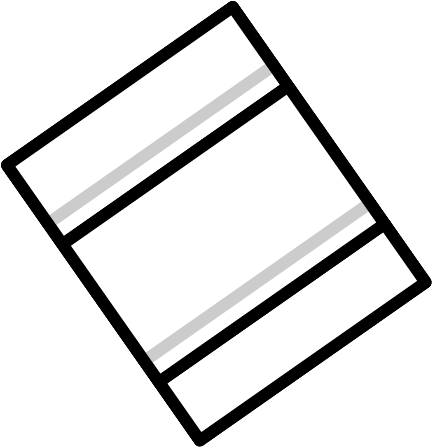} & \includegraphics[scale=0.1,align=c]{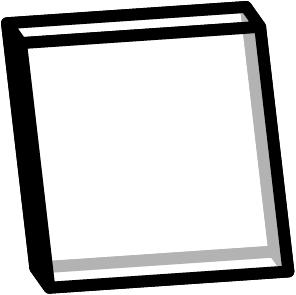} \\[5pt] 
 & $-$ & & $-$ &\num{90.00000000000001} & \includegraphics[scale=0.1,align=c]{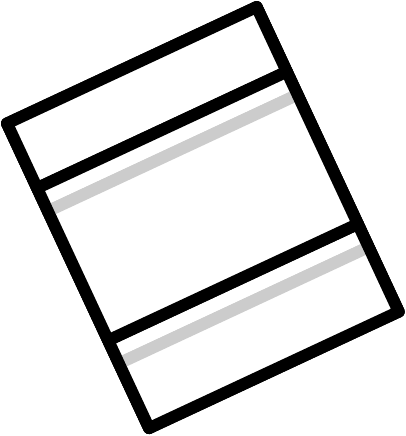} & \includegraphics[scale=0.1,align=c]{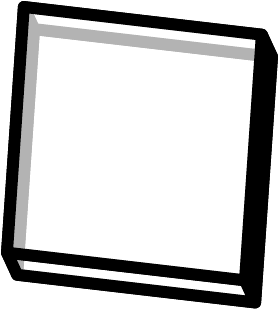} \\[5pt] 
 & $-$ & & $+$ &\num{90.00000000000001} & \includegraphics[scale=0.1,align=c]{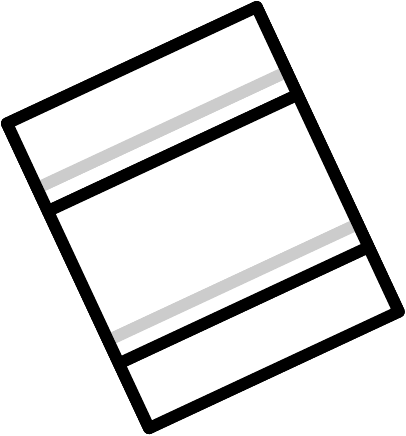} & \includegraphics[scale=0.1,align=c]{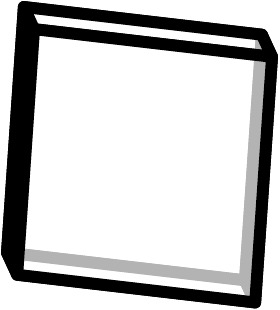} \\[5pt] 

 \hline 
 & $+$ & & $-$ &\num{60.00000000000001} & \includegraphics[scale=0.1,align=c]{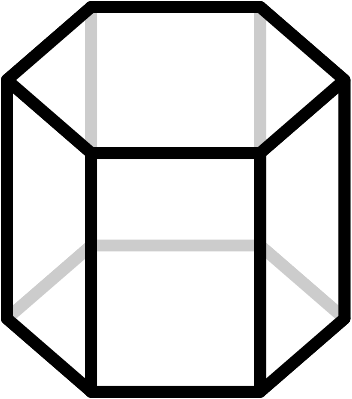} & \includegraphics[scale=0.1,align=c]{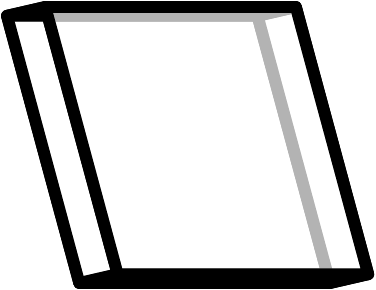} \\[5pt] 
\multirow{2}*{$\text{h}_{2}$} & $+$ & \multirow{2}*{ $\text{b}_{3}$ } & $+$ &\num{60.83197478497548} & \includegraphics[scale=0.1,align=c]{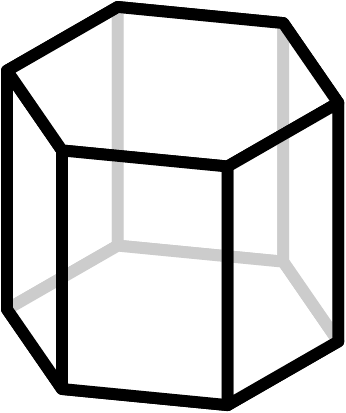} & \includegraphics[scale=0.1,align=c]{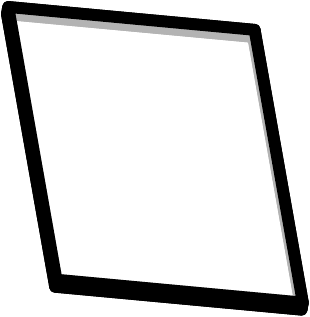} \\[5pt] 
 & $-$ & & $-$ &\num{60.83197478497548} & \includegraphics[scale=0.1,align=c]{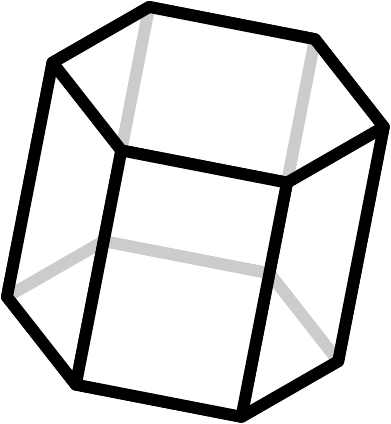} & \includegraphics[scale=0.1,align=c]{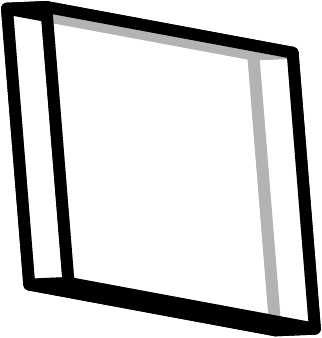} \\[5pt] 
 & $-$ & & $+$ &\num{63.26177218039435} & \includegraphics[scale=0.1,align=c]{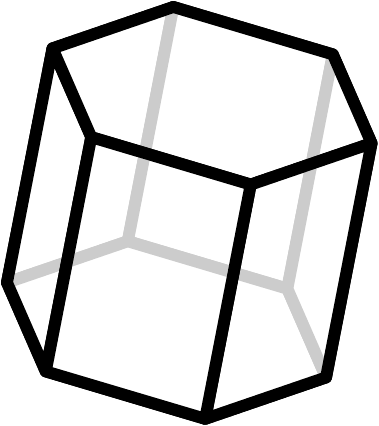} & \includegraphics[scale=0.1,align=c]{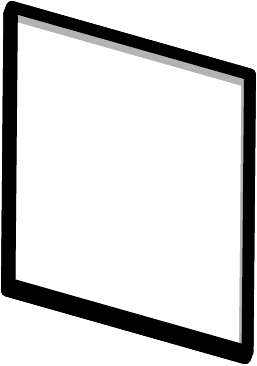} \\[5pt] 

 \hline 
 & $+$ & & $-$ &\num{63.26177218039435} & \includegraphics[scale=0.1,align=c]{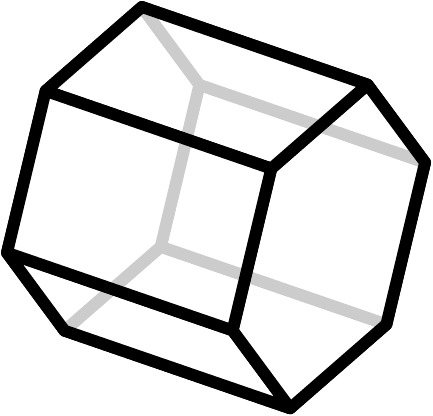} & \includegraphics[scale=0.1,align=c]{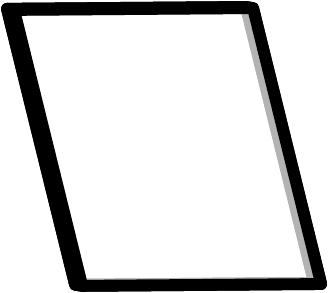} \\[5pt] 
\multirow{2}*{$\text{h}_{2}$} & $+$ & \multirow{2}*{ $\text{b}_{4}$ } & $+$ &\num{60.83197478497545} & \includegraphics[scale=0.1,align=c]{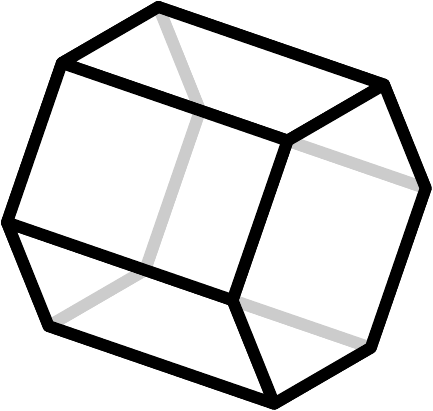} & \includegraphics[scale=0.1,align=c]{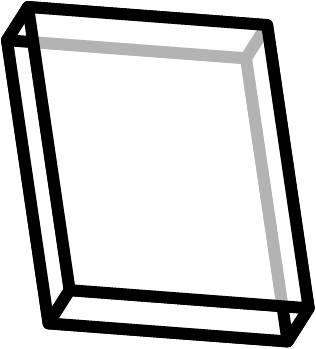} \\[5pt] 
 & $-$ & & $-$ &\num{60.83197478497545} & \includegraphics[scale=0.1,align=c]{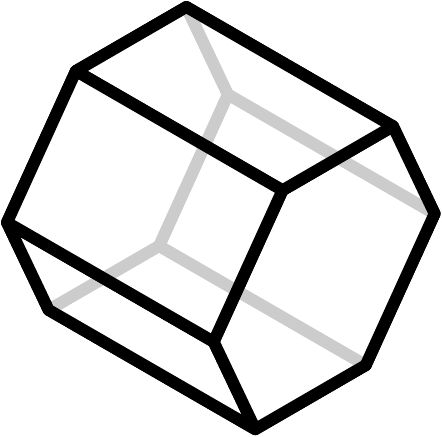} & \includegraphics[scale=0.1,align=c]{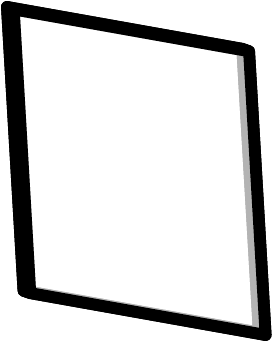} \\[5pt] 

 \hline 
 & $+$ & & $-$ &\num{60.00000000000001} & \includegraphics[scale=0.1,align=c]{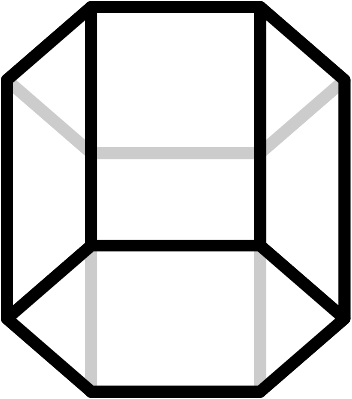} & \includegraphics[scale=0.1,align=c]{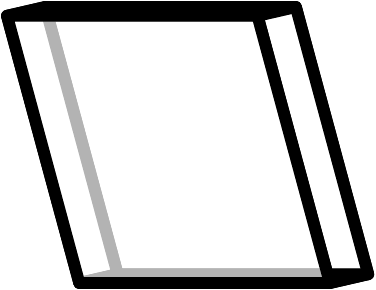} \\[5pt] 
\multirow{2}*{$\text{h}_{2}$} & $+$ & \multirow{2}*{ $\text{b}_{9}$ } & $+$ &\num{60.83197478497548} & \includegraphics[scale=0.1,align=c]{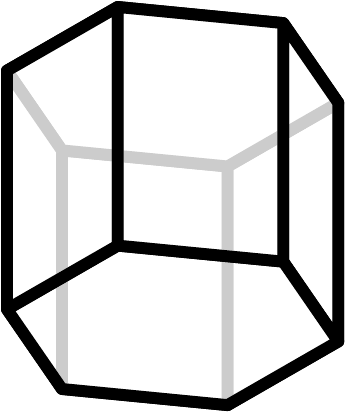} & \includegraphics[scale=0.1,align=c]{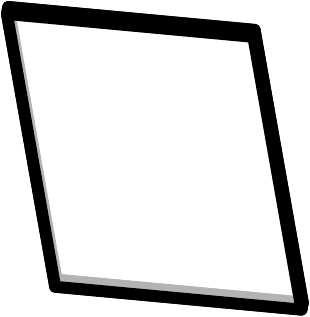} \\[5pt] 
 & $-$ & & $-$ &\num{60.83197478497548} & \includegraphics[scale=0.1,align=c]{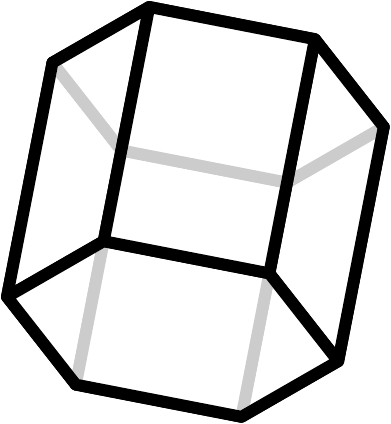} & \includegraphics[scale=0.1,align=c]{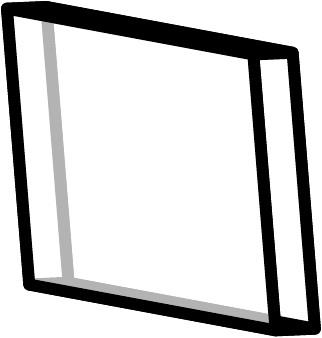} \\[5pt] 
 & $-$ & & $+$ &\num{63.26177218039435} & \includegraphics[scale=0.1,align=c]{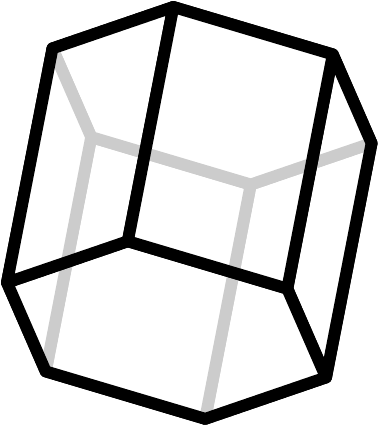} & \includegraphics[scale=0.1,align=c]{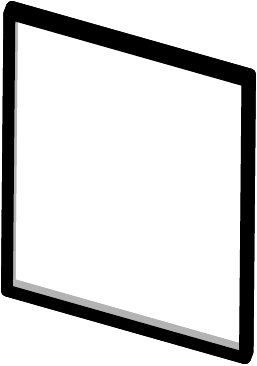} \\[5pt] 

 \hline 
 & $+$ & & $-$ &\num{63.261772180394324} & \includegraphics[scale=0.1,align=c]{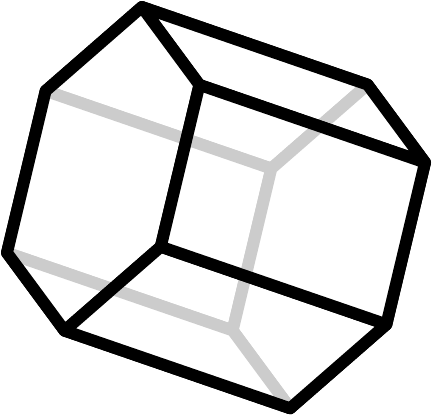} & \includegraphics[scale=0.1,align=c]{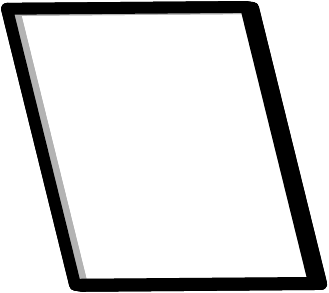} \\[5pt] 
\multirow{2}*{$\text{h}_{2}$} & $+$ & \multirow{2}*{ $\text{b}_{10}$ } & $+$ &\num{60.83197478497545} & \includegraphics[scale=0.1,align=c]{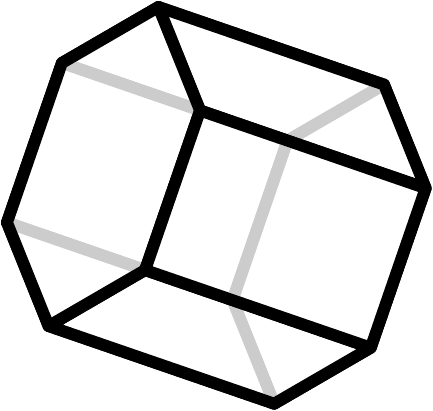} & \includegraphics[scale=0.1,align=c]{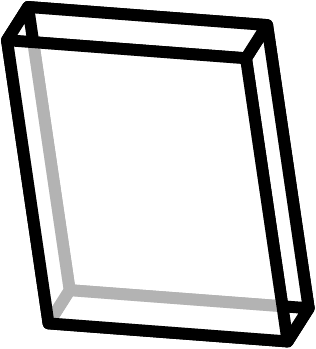} \\[5pt] 
 & $-$ & & $-$ &\num{60.83197478497545} & \includegraphics[scale=0.1,align=c]{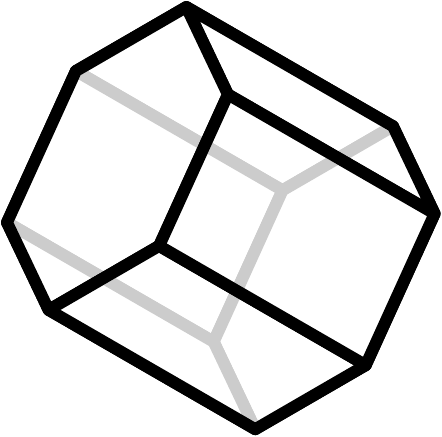} & \includegraphics[scale=0.1,align=c]{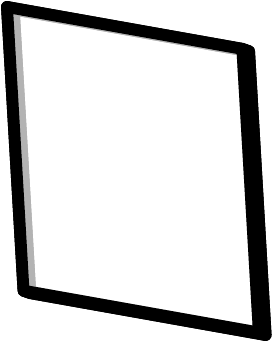} \\[5pt] 

 \hline 
 & $+$ & & $-$ &\num{90.00000000000001} & \includegraphics[scale=0.1,align=c]{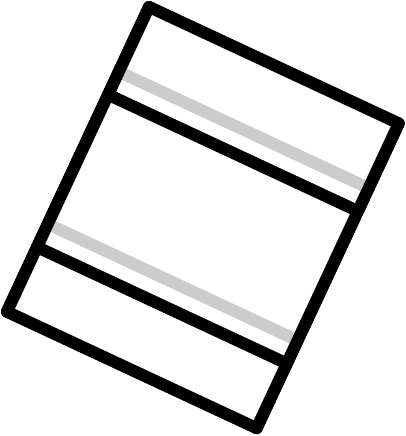} & \includegraphics[scale=0.1,align=c]{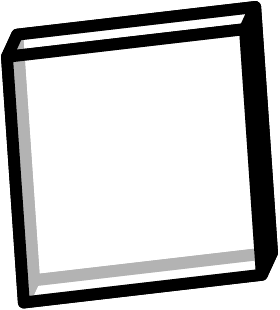} \\[5pt] 
\multirow{2}*{$\text{h}_{3}$} & $+$ & \multirow{2}*{ $\text{b}_{2}$ } & $+$ &\num{90.00000000000001} & \includegraphics[scale=0.1,align=c]{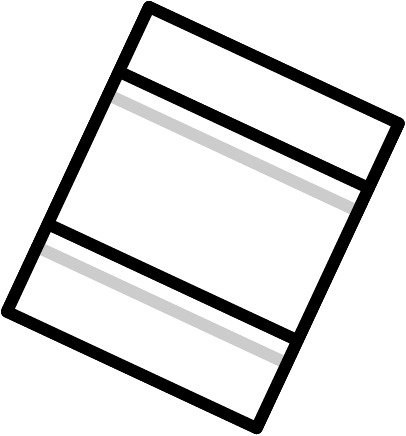} & \includegraphics[scale=0.1,align=c]{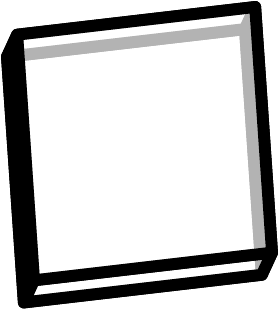} \\[5pt] 
 & $-$ & & $-$ &\num{90.00000000000001} & \includegraphics[scale=0.1,align=c]{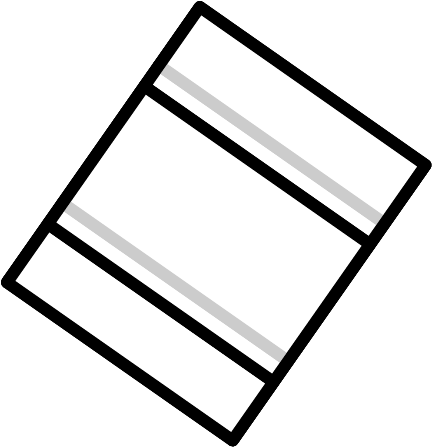} & \includegraphics[scale=0.1,align=c]{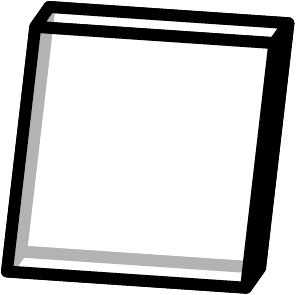} \\[5pt] 
 & $-$ & & $+$ &\num{90.00000000000001} & \includegraphics[scale=0.1,align=c]{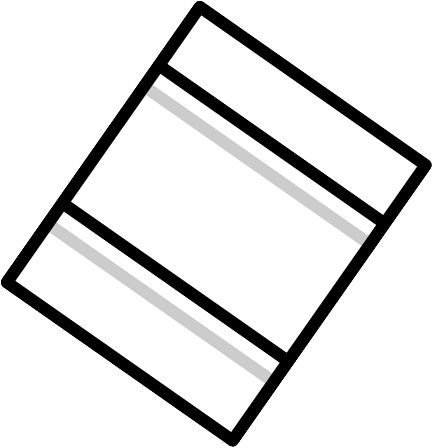} & \includegraphics[scale=0.1,align=c]{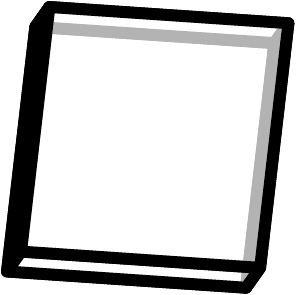} \\[5pt] 

 \hline 
 & $+$ & & $+$ &\num{60.83197478497545} & \includegraphics[scale=0.1,align=c]{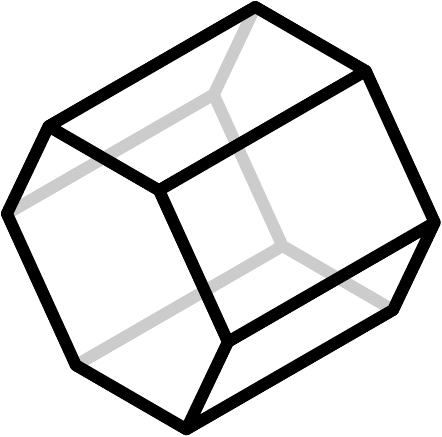} & \includegraphics[scale=0.1,align=c]{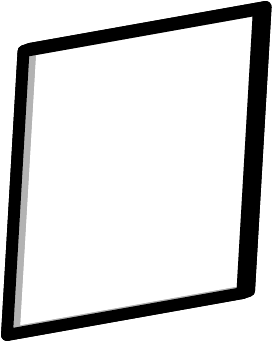} \\[5pt] 
\multirow{2}*{$\text{h}_{3}$} & $-$ & \multirow{2}*{ $\text{b}_{3}$ } & $-$ &\num{60.83197478497545} & \includegraphics[scale=0.1,align=c]{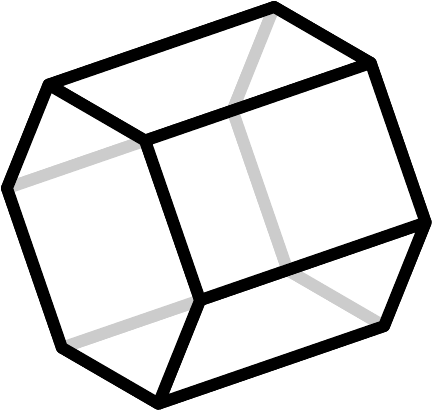} & \includegraphics[scale=0.1,align=c]{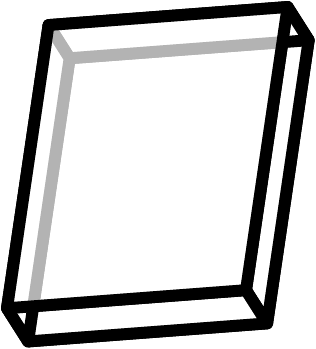} \\[5pt] 
 & $-$ & & $+$ &\num{63.26177218039435} & \includegraphics[scale=0.1,align=c]{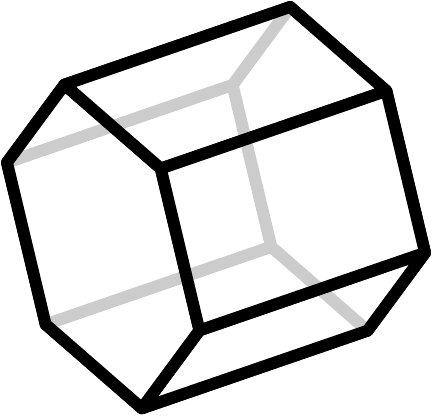} & \includegraphics[scale=0.1,align=c]{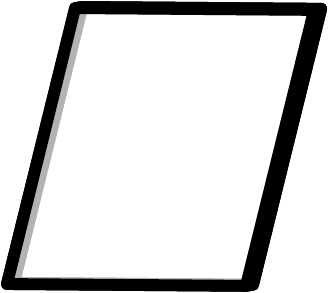} \\[5pt] 

 \hline 
 & $+$ & & $-$ &\num{63.26177218039435} & \includegraphics[scale=0.1,align=c]{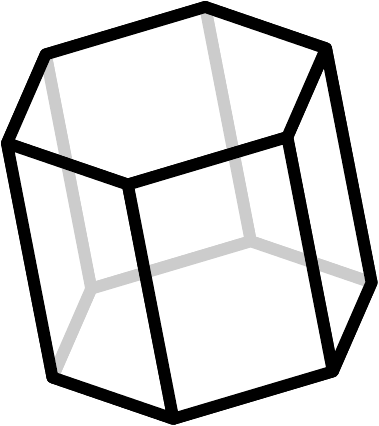} & \includegraphics[scale=0.1,align=c]{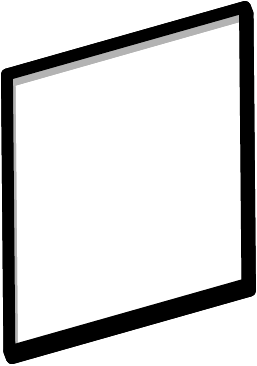} \\[5pt] 
\multirow{2}*{$\text{h}_{3}$} & $+$ & \multirow{2}*{ $\text{b}_{4}$ } & $+$ &\num{60.83197478497548} & \includegraphics[scale=0.1,align=c]{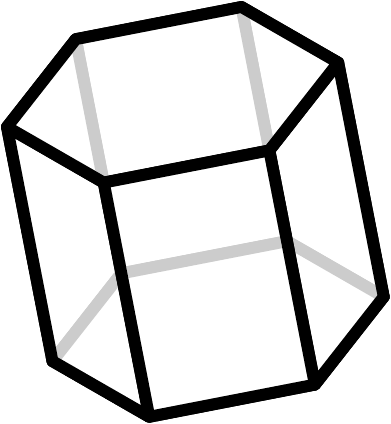} & \includegraphics[scale=0.1,align=c]{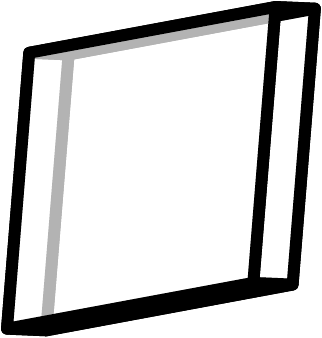} \\[5pt] 
 & $-$ & & $-$ &\num{60.83197478497548} & \includegraphics[scale=0.1,align=c]{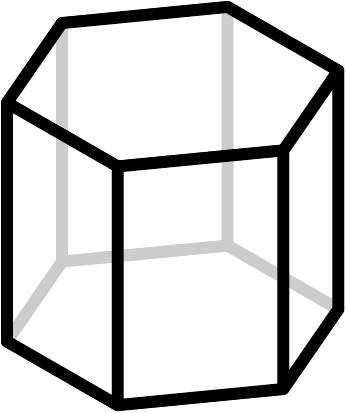} & \includegraphics[scale=0.1,align=c]{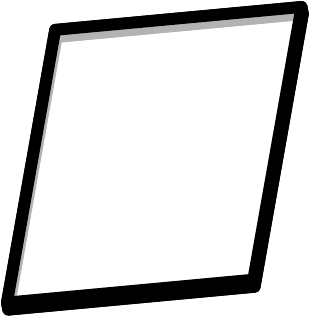} \\[5pt] 

 \hline 
 & $+$ & & $+$ &\num{60.83197478497545} & \includegraphics[scale=0.1,align=c]{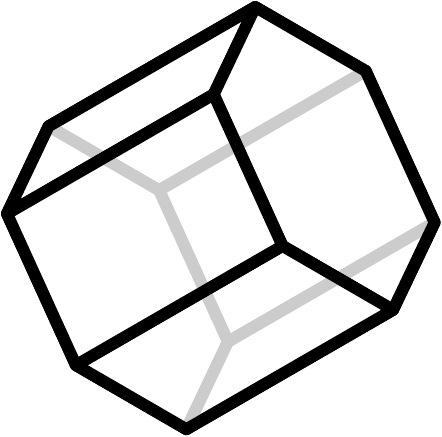} & \includegraphics[scale=0.1,align=c]{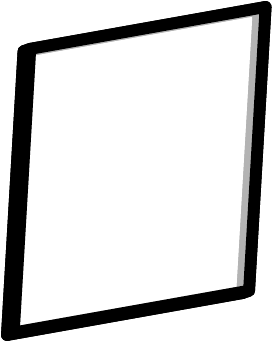} \\[5pt] 
\multirow{2}*{$\text{h}_{3}$} & $-$ & \multirow{2}*{ $\text{b}_{9}$ } & $-$ &\num{60.83197478497545} & \includegraphics[scale=0.1,align=c]{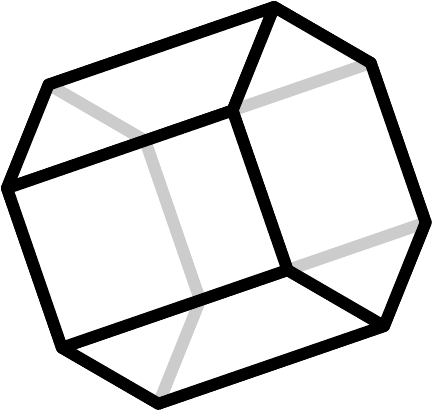} & \includegraphics[scale=0.1,align=c]{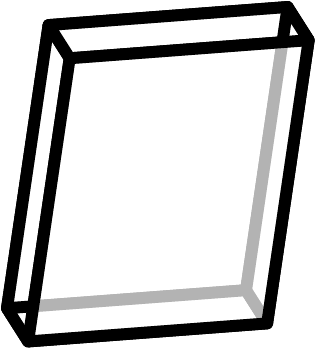} \\[5pt] 
 & $-$ & & $+$ &\num{63.261772180394324} & \includegraphics[scale=0.1,align=c]{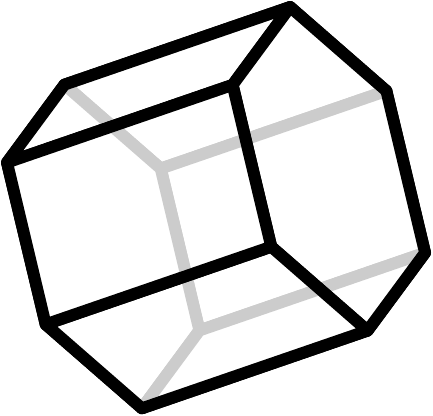} & \includegraphics[scale=0.1,align=c]{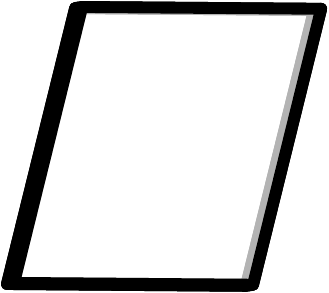} \\[5pt] 

 \hline 
 & $+$ & & $-$ &\num{63.26177218039435} & \includegraphics[scale=0.1,align=c]{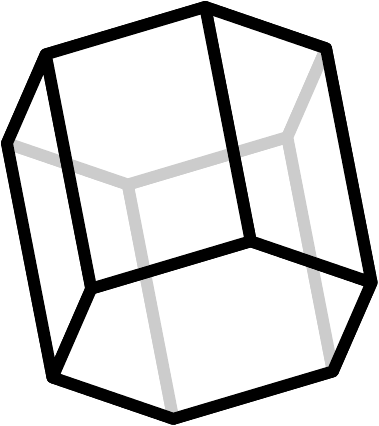} & \includegraphics[scale=0.1,align=c]{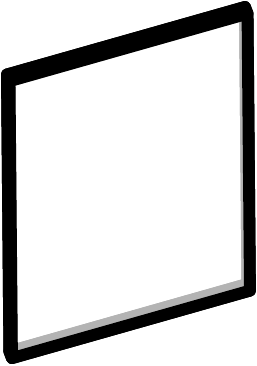} \\[5pt] 
\multirow{2}*{$\text{h}_{3}$} & $+$ & \multirow{2}*{ $\text{b}_{10}$ } & $+$ &\num{60.83197478497548} & \includegraphics[scale=0.1,align=c]{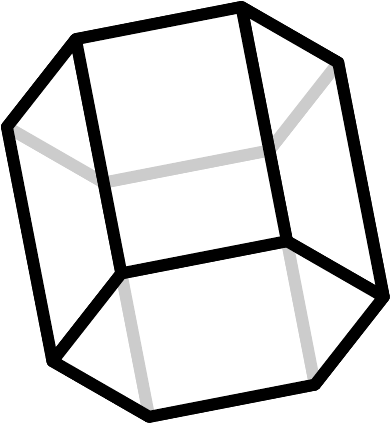} & \includegraphics[scale=0.1,align=c]{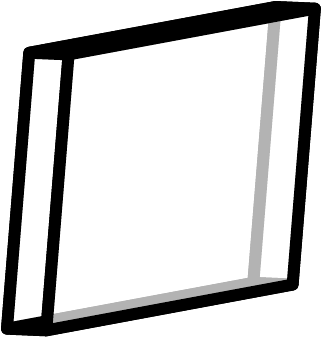} \\[5pt] 
 & $-$ & & $-$ &\num{60.83197478497548} & \includegraphics[scale=0.1,align=c]{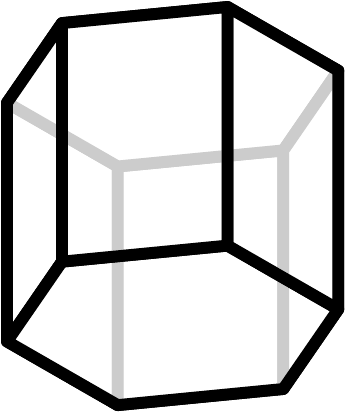} & \includegraphics[scale=0.1,align=c]{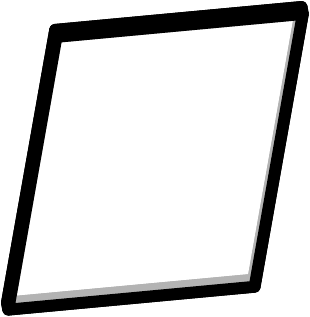} \\[5pt] 

 \bottomrule
 
 \end{xtabular} 
\end{center}

\begin{table}[h!]
\caption{List of 24 cubic lattice symmetry operators used in the thesis.}
\label{tab:symbcc}
\begin{center}
%\begin{tabular}{ M{4cm} M{4cm} M{4cm} } 
\begin{tabular}{ cccccccc } 
\toprule
 $\text{b}_{1}$ & 
 $\begin{bmatrix} 1& 0&0 \\ 0 & 1& 0 \\ 0 & 0 & 1 \end{bmatrix}$ &
 $\text{b}_{2}$ & 
 $\begin{bmatrix} -1&0 &0 \\ 0&-1& 0 \\ 0&0&1 \end{bmatrix}$ &
 $\text{b}_{3}$ &
 $\begin{bmatrix} 0&-1&0 \\ -1&0& 0 \\ 0&0&-1 \end{bmatrix}$ &
 $\text{b}_{4}$ &
 $\begin{bmatrix} 0&1&0 \\ 1&0& 0 \\ 0&0&-1 \end{bmatrix}$\\[1cm] 

 $\text{b}_{5}$ &
  $\begin{bmatrix} 1&0&0 \\ 0&-1& 0 \\ 0&0&-1 \end{bmatrix}$ &
 $\text{b}_{6}$ &
  $\begin{bmatrix} -1&0&0 \\ 0&1& 0 \\ 0&0&-1 \end{bmatrix}$ &
 $\text{b}_{7}$ &
  $\begin{bmatrix} 0&1&0 \\ -1&0& 0 \\ 0&0&1 \end{bmatrix}$ &
 $\text{b}_{8}$ &
  $\begin{bmatrix} 0&-1&0 \\ 1&0& 0 \\ 0&0&1 \end{bmatrix}$\\[1cm] 

 $\text{b}_{9}$ &
  $\begin{bmatrix} 0&0&-1 \\ 1&0& 0 \\ 0&-1&0 \end{bmatrix}$ &
 $\text{b}_{10}$ & 
  $\begin{bmatrix} 0&0&1 \\ -1&0& 0 \\ 0&-1&0 \end{bmatrix}$ &
 $\text{b}_{11}$ & 
  $\begin{bmatrix} 1&0&0 \\ 0&0&-1 \\ 0&1&0 \end{bmatrix}$ &
 $\text{b}_{12}$ &  
  $\begin{bmatrix} -1&0&0 \\ 0&0&1 \\ 0&1&0 \end{bmatrix}$\\[1cm] 

 $\text{b}_{13}$ & 
  $\begin{bmatrix} 0&0&-1 \\ -1&0& 0 \\ 0&1&0 \end{bmatrix}$ &
 $\text{b}_{14}$ & 
  $\begin{bmatrix} 0&0&1 \\ 1&0& 0 \\ 0&1&0 \end{bmatrix}$ &
 $\text{b}_{15}$ & 
  $\begin{bmatrix} -1&0&0 \\ 0&0&-1 \\ 0&-1&0 \end{bmatrix}$ &
$\text{b}_{16}$ &  
 $\begin{bmatrix} 1&0&0 \\ 0&0&1 \\ 0&-1&0 \end{bmatrix}$\\[1cm] 

 $\text{b}_{17}$ & 
  $\begin{bmatrix} 0&1&0 \\ 0&0&1 \\ 1&0&0 \end{bmatrix}$ &
$\text{b}_{18}$ &  
 $\begin{bmatrix} 0&-1&0 \\ 0&0&-1 \\ 1&0&0 \end{bmatrix}$ &
$\text{b}_{19}$ & 
 $\begin{bmatrix} 0&0&-1 \\ 0&-1& 0 \\ -1&0&0 \end{bmatrix}$ &
 $\text{b}_{20}$ &
  $\begin{bmatrix} 0&0&1 \\ 0&1& 0 \\ -1&0&0 \end{bmatrix}$\\[1cm] 

 $\text{b}_{21}$ & 
  $\begin{bmatrix} 0&1&0 \\ 0&0&-1 \\ -1&0&0 \end{bmatrix}$ &
$\text{b}_{22}$ &  
 $\begin{bmatrix} 0&-1&0 \\ 0&0&-1 \\ 1&0&0 \end{bmatrix}$ &
 $\text{b}_{23}$ & 
  $\begin{bmatrix} 0&0&1 \\ 0&-1& 0 \\ 1&0&0 \end{bmatrix}$ &
$\text{b}_{24}$ &  
 $\begin{bmatrix} 0&0&-1 \\ 0&1& 0 \\ 1&0&0 \end{bmatrix}$\\[1cm]  
\bottomrule
\end{tabular}
\end{center}
\end{table}

%\end{linenumbers}
\end{document}